\documentclass[sigconf]{acmart}

\usepackage{amsmath}
\usepackage{algorithm}
\usepackage{algpseudocode}
\usepackage{algcompatible}
\algnewcommand\algorithmicreturn{\textbf{return}}
\algnewcommand\RETURN{\State \algorithmicreturn}
\usepackage{amsmath}
\usepackage{geometry}
\usepackage{algpseudocode}
\usepackage{graphicx}
\usepackage{subfigure}
\usepackage{textcomp}
\usepackage{xcolor}
\usepackage{booktabs}
\usepackage{multirow}
\usepackage{array}
\usepackage{caption}
\usepackage{float}
\usepackage{balance}

\newcommand{\score}{\textit{humanoid}}
\newcommand{\XR}{Metaverse}
\newcommand{\problem}{MetaCD}

\copyrightyear{2025}
\acmYear{2025}
\setcopyright{acmlicensed}\acmConference[WWW '25]{Proceedings of the ACM Web Conference 2025}{April 28-May 2, 2025}{Sydney, NSW, Australia}
\acmBooktitle{Proceedings of the ACM Web Conference 2025 (WWW '25), April 28-May 2, 2025, Sydney, NSW, Australia}
\acmDOI{10.1145/3696410.3714679}
\acmISBN{979-8-4007-1274-6/25/04}

\begin{document}
\title{Human-Centric Community Detection in Hybrid Metaverse Networks with Integrated AI Entities}

\author{Shih-Hsuan Chiu}
\email{shchiu@arbor.ee.ntu.edu.tw}
\affiliation{%
  \institution{National Taiwan University}
  \city{Taipei}
  \country{Taiwan}
}

\author{Ya-Wen Teng}
\email{ywteng@iis.sinica.edu.tw}
\affiliation{%
  \institution{Academia Sinica}
  \city{Taipei}
  \country{Taiwan}
}

\author{De-Nian Yang}
\email{dnyang@iis.sinica.edu.tw}
\affiliation{%
  \institution{Academia Sinica}
  \city{Taipei}
  \country{Taiwan}
}

\author{Ming-Syan Chen}
\email{mschen@ntu.edu.tw}
\affiliation{%
  \institution{National Taiwan University}
  \city{Taipei}
  \country{Taiwan}
}

\begin{abstract}
Community detection is a cornerstone problem in social network analysis (SNA), aimed at identifying cohesive communities with minimal external links. However, the rise of generative AI and Metaverse introduce complexities by creating hybrid human-AI social networks (denoted by HASNs), where traditional methods fall short, especially in human-centric settings. This paper introduces a novel community detection problem in HASNs (denoted by MetaCD), which seeks to enhance human connectivity within communities while reducing the presence of AI nodes. Effective processing of MetaCD poses challenges due to the delicate trade-off between excluding certain AI nodes and maintaining community structure. To address this, we propose CUSA, an innovative framework incorporating AI-aware clustering techniques that navigate this trade-off by selectively retaining AI nodes that contribute to community integrity. Furthermore, given the scarcity of real-world HASNs, we devise four strategies for synthesizing these networks under various hypothetical scenarios. Empirical evaluations on real social networks, reconfigured as HASNs, demonstrate the effectiveness and practicality of our approach compared to traditional non-deep learning and graph neural network (GNN)-based methods.
\end{abstract}



\maketitle

\begin{figure}[t]
    \centering
    \includegraphics[width=0.76\linewidth]{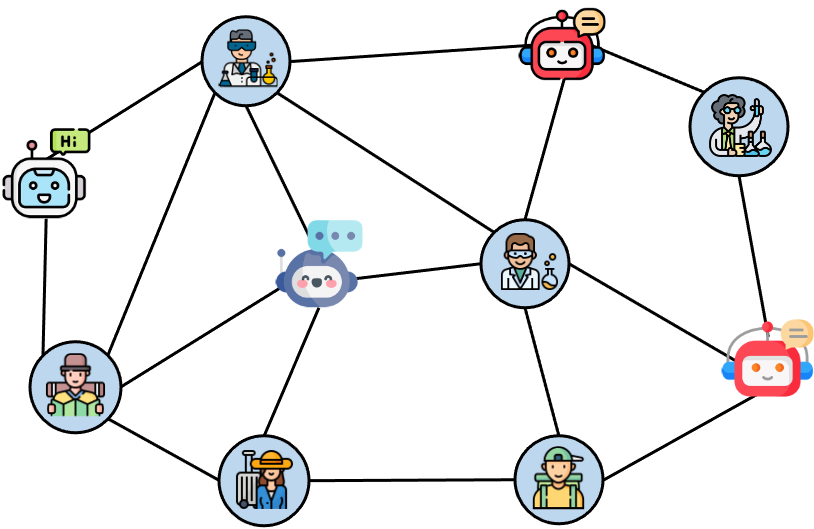}
    \caption{A schematic depiction of HASN, illustrating the interconnected relationships between human users and AI entities. This paper seeks to identify \textit{human-centric communities} within this hybrid network.}
    \label{figure:hasn}
\end{figure}

\section{Introduction}
\label{sec:introduction}

Community detection is a fundamental problem in social network analysis (SNA), focusing on identifying tightly-knit communities with minimal external connections \cite{mcpherson2001birds}\cite{su2022comprehensive}\cite{jin2021survey}. This task is critical for analyzing social relationships and holds practical applications in marketing and personalized services, where insights into community structures enable targeted strategies and enhance user engagement \cite{du2007community}\cite{umrawal2023community}.

The integration of generative AI and Metaverse is transforming traditional social networks, creating human-AI social networks (HASNs) (Figure \ref{figure:hasn}) that blend human and AI entities within shared digital spaces. In Metaverse environments, which combine virtual reality (VR) and mixed reality (MR) components, users engage with both humans and AI-driven entities, such as avatars and virtual assistants, forming communities that span the physical and digital worlds. For example, social Metaverse platforms like Microsoft AltspaceVR \cite{meijers2021globalxr} and Meta Horizon Workrooms \cite{meta2023workrooms} enable immersive, large-scale virtual gatherings, allowing users to interact with AI alongside human participants through customizable 3D environments, spatial audio, and interactive tools.

\begin{figure*}[t]
    \centering
    \includegraphics[height=0.2275\linewidth]{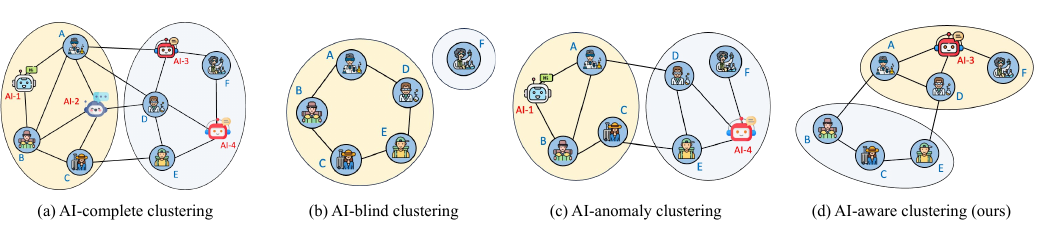}
    \caption{Illustrative examples of community detection in an HASN.}
    \label{illustrative}
\end{figure*}

In these evolving HASNs, a fundamental shift occurs: traditional social clusters now include AI nodes, making human-centric community detection crucial. Unlike conventional networks, HASNs often aim to prioritize human interactions while accommodating AI entities that provide social support or assistance. For instance, platforms such as Nomi AI \cite{nomiAI2024} create AI companions for users, enhancing social engagement and addressing issues like loneliness. Similarly, platforms like Engage \cite{engage2023} and Mozilla Hubs \cite{hubs2023mozilla} host virtual events and social gatherings, integrating AI to foster rich, interactive experiences. These hybrid networks demand new community detection methods that discover human-centric communities by selectively retaining AI nodes that strengthen community cohesion and removing those that do not. This approach supports applications focused on authentic human engagement, including marketing, user recommendations, and digital companionship.

However, applying traditional community detection methods to these hybrid networks presents several challenges. For instance, (1) methods that ignore AI entities may result in communities with an excess of AIs, which are less effective for human-centric applications such as advertising and recommendation, as these efforts are not relevant for AI entities. Moreover, this approach can also lead to incorrect community assignments for human members. (2) Alternatively, one might consider removing all AI nodes before conducting community detection. However, this could disrupt connections between AI and human nodes, as well as among AI nodes themselves. Since community detection relies heavily on graph topology, these disruptions may lead to inaccurate community assignments for human users. (3) A third approach could involve using anomaly detection techniques to classify certain AI nodes as anomalies and remove them before clustering. However, AI node behavior often diverges from conventional “anomalies” in traditional social networks \cite{ma2021comprehensive}\cite{lamichhane2024anomaly}. Unlike typical outliers, AI nodes may mimic human behavior, support human interactions, and integrate into communities, making this strategy likely to yield unsatisfactory results.

In this work, we introduce a novel community detection problem tailored to scenarios where humans and numerous AI entities are intertwined within social networks\footnote{To the best of our knowledge, this is the first study to explore community detection in the hybrid human-AI scenario. While focusing on structural information, our approach is designed to be compatible with future integration of semantic features, enabling more refined human-centric clustering based on shared interests or interaction patterns.}, emphasizing human closeness derived from the graph structure. We envision an \XR\ environment where a social network comprises both human users and AI entities, forming what we term the Human-AI Social Network (HASN), as illustrated in Figure \ref{figure:hasn}. Accordingly, we envisage four scenarios of HASNs existing in \XR\ \cite{techcrunch_meta_2024}: (1) Random Interaction, (2) Introverted Humans Prioritized for Interaction, (3) Distinct AI Types, and (4) Dual-personality AIs. These proposed scenarios are grounded in the versatility of AI, allowing it to interact widely and adaptively with diverse users \cite{wang2024survey}. (Details are provided in the experimental setup, Section \ref{subsec:experimental_setup}.)

This work focuses on community detection in HASNs, especially when the primary focus is on human users (human-centric communities). Ideally, clustering should produce clusters with high human closeness and minimal AI presence, balancing the removal of AI nodes while maintaining community integrity. Let us explore some illustrative examples as follows. Figure \ref{illustrative} shows several approaches to cluster an HASN depending on how they consider the AI nodes: (1) \textit{AI-complete clustering}: Figure \ref{illustrative} (a) shows a clustering result from a method that ignores AI nodes and performs clustering directly. However, as observed, each community contains an excessive number of AI entities, making it impractical for human-centric applications such as advertising and recommendation. In addition, this approach may result in inaccurate community assignments for human members (e.g., A and D, C and E). 
(2) \textit{AI-blind clustering}: Figure \ref{illustrative} (b) shows a clustering result from a method that treats all AI nodes as outliers and removes them before clustering. Human nodes A and D are initially connected to node F through AI node AI-3, suggesting that A, D, and F may belong to the same community. However, this approach can disrupt connections between AI and human nodes, as well as among AI nodes, resulting in the loss of important links and leading to unsatisfactory community results. (3) \textit{AI-anomaly clustering}: Figure \ref{illustrative} (c) shows a clustering result where certain AI nodes are identified as anomalies through anomaly detection techniques and removed before clustering. However, unlike typical anomalies in traditional social networks, which often link different communities or form dense connections \cite{ma2021comprehensive}\cite{lamichhane2024anomaly}, AI nodes may mimic human behavior, support human interactions, and integrate into communities. Consequently, as shown, traditional anomaly detection would remove AI-2 and AI-3, resulting in inaccurate community outcomes for human users (e.g., A and F). (4) \textit{AI-aware clustering}: Figure \ref{illustrative} (d) illustrates clustering results achieved by identifying AI nodes that serve as bridges between humans or connect key human nodes. For instance, AI-3 is an AI node retained during the clustering process because it effectively links humans A, D, and F. This connectivity does not occur in the aforementioned three methods (i.e., Figure \ref{illustrative} (a), (b), and (c)). In other words, this approach preserves AI nodes that positively impact the community while removing those that do not, thereby enhancing human closeness and fostering potential human-centric communities.

Considering the factors depicted above face the following challenges: (i) \textit{Evaluation of AI nodes}: The behavior patterns of AI nodes may not resemble the typical ”anomalies” found in traditional social networks \cite{ma2021comprehensive}. AI may mimic human behavior, assist humans, and integrate into communities. In other words, some AI nodes are helpful for the formation of potential communities, while some are redundant. Accordingly, traditional anomaly detection methods are inadequate for evaluating AI nodes in emerging HASN graphs. (ii) \textit{Tradeoff between AI removal and community integrity}: Removing AI nodes could lead to disconnections between AI and humans, as well as among AIs themselves. Therefore, it is crucial to identify and preserve AI nodes that enhance human closeness while removing those that do not contribute positively. For instance, some AI nodes act as bridges, enhancing human connections and facilitating the formation of potential communities. (iii) \textit{Avoidance of local optima in AI-aware clustering}: The search space for finding the optimal combination of AI preservation/removal expands exponentially as the number of AI increases. This complexity makes this problem hard to solve and may result in solutions that settle at the local optima.

In this paper, we formulate a new problem, termed \textbf{\textit{human-centric community detection in hybrid networks (HASNs) of \XR}} (denoted by \problem). Unlike previous studies focusing on community detection in purely human networks without considering AI involvement \cite{su2022comprehensive}\cite{jin2021survey}, our approach explores community detection within hybrid networks, such as HASNs (illustrated in Figure \ref{figure:hasn}), composed of both human users and AI entities.  Given an HASN with prior knowledge of which nodes in the network are AI nodes, our goal is to generate clusters that can \textit{maximize human closeness with minimal AI involvement}, by balancing the removal of AI nodes and maintaining community integrity. Specifically, a desirable clustering result of an HASN should achieve two key objectives simultaneously: (1) maximizing human closeness and (2) minimizing the presence of AI nodes within each cluster \footnote{For more information on potential application scenarios of MetaCD, please refer to Appendix \ref{sec: Potential Application Scenarios}.}.

To this end, we design a novel algorithm called \textit{\underline{Cu}stomized AI-aware \underline{S}imulated \underline{A}nnealing for Clustering} (denoted by CUSA), to tackle the above challenges of \problem. \textbf{(1)} For the first challenge in the evaluation of AI nodes, CUSA incorporates \textit{AI Scoring} to evaluate the "\score" of an AI node. Specifically, if an AI node acts to a greater extent as a bridge between humans or connects important human nodes, its \score\ is higher, and vice versa. This is because it can enhance human closeness and potentially lead to the formation of new human communities or the migration of humans to different communities due to the influence of such AI nodes. \textbf{(2)} For the second challenge in the tradeoff between AI removal and community integrity, we develop the \textit{AI-aware Louvain clustering algorithm} into CUSA. This algorithm adaptively groups nodes based on human closeness gain while also taking into account the proportion of AI presence in each community during clustering. \textbf{(3)} For the third challenge in avoidance of local optima in AI-aware clustering, CUSA infuses \textit{AI-aware adaptive clustering (3AC) framework with a probability-based escape strategy}. This framework adaptively partitions an HASN by removing the AI node with the lowest \score. In addition to removing (or preserving) AI nodes based on their \score, we employ a pre-defined probability distribution to escape local optima. We evaluate the performance of CUSA on benchmark real-world social networks (i.e., Cora,  CiteSeer, and PubMed) transformed into HASNs (Figure \ref{figure:hasn}) with our proposed generation strategies. The contributions of this work include:

\begin{itemize}
    \item To the best of our knowledge, \problem\ is the first attempt to study the community detection problem under the new scenarios where humans and numerous AI entities are intertwined in social networks (denoted by HASNs), especially focusing on human-centric communities. In addition, we propose four HASN scenarios, each with specifically designed generation strategies.
    
    \item To effectively address \problem, we develop a novel algorithm called CUSA, incorporating AI-aware clustering techniques that navigate the delicate trade-off between removing AI nodes and maintaining community structure by selectively retaining AI nodes that contribute to community integrity. 
    
    \item Empirical evaluations on real social networks (reconfigured as HASNs) demonstrate the effectiveness of CUSA in forming human-centric communities compared to competitive baselines.

    \item The experiments show that carefully designed generation strategies can improve clustering results, enhancing human closeness and revealing potential communities.
    
\end{itemize}


\section{Related work}
\label{sec:related_work}

\noindent\textbf{Community Detection.} Community detection is the process of grouping nodes into densely connected communities with sparse interconnections, depending on the structure of the graph. Early traditional non-deep learning methods, such as the spectral clustering algorithm \cite{amini2013pseudo}, optimize ratio and normalized-cut criteria. Louvain \cite{blondel2008fast} is a well-known optimization algorithm that uses a node-moving strategy to optimize modularity. Extensions of greedy optimizations include simulated annealing \cite{kirkpatrick1983optimization}, extremal optimization \cite{boettcher2002optimization}, and spectral optimization \cite{newman2013spectral}.Recently, deep learning-based community detection techniques \cite{su2022comprehensive}\cite{lecun2015deep} have gained traction, with graph neural networks (GNNs) emerging as a key trend. These methods learn lower-dimensional representations from high-dimensional structural data \cite{khoshraftar2024survey} \cite{tsitsulin2023graph}, capturing diverse information from nodes \cite{ge2024unsupervised}, edges \cite{chikwendu2023comprehensive}, neighborhoods \cite{wang2023overview}, or multigraphs \cite{pan2021multi}. Moreover, deep learning effectively uncovers community structures in large-scale \cite{fang2020survey}, sparse \cite{wu2020deep}, complex \cite{shao2024distributed}, and dynamic networks \cite{pan2024identification}. Despite the extensive exploration of various community detection methods and their application in different real-world scenarios, most existing approaches generally operate under the common assumption that social networks are solely composed of human users. In this work, we investigate a novel community detection problem tailored for social networks where humans and numerous AI entities coexist and interact. To address this, we propose a new method that prioritizes human-centric communities while effectively addressing the complexities introduced by AI entities within human-AI social networks (HASNs).

\noindent\textbf{Graph Anomaly Detection.} Graph Anomaly Detection (GAD) is designed to identify unusual patterns, outliers, or unexpected behaviors in graph-structured data \cite{ma2021comprehensive}\cite{lamichhane2024anomaly}. GAD techniques have proven effective in various applications, including computer network intrusion detection \cite{bilot2023graph}, fraud detection \cite{xiang2023semi}, and anomaly detection in social networks \cite{yu2016survey}\cite{roy2024gad}. In contrast, our work addresses a novel issue by examining interactions between humans and AI in a newly defined social network (HASN). Unlike traditional social networks, where “anomalies” are typically detected through unusual connection patterns—such as bridging distinct communities or establishing dense links with other nodes \cite{ma2021comprehensive}—AI behavior may not inherently appear abnormal. AI can mimic human behavior, assist humans, and seamlessly integrate into communities. Consequently, this distinction makes existing GAD methods ill-suited for addressing our new problem.

\noindent\textbf{Generative Artificial Intelligence.} Generative Artificial Intelligence (GAI) is a form of AI capable of autonomously generating new content, including text, images, audio, and video. The current mainstream approach to realizing GAI involves training large language models (LLMs) \cite{guo2024large}. Various applications of LLMs have rapidly emerged, including ChatGPT \cite{achiam2023gpt}, Gemini \cite{team2023gemini}, and Claude \cite{AnthropicAI2023}. These LLM applications have significantly transformed our lives by adding convenience in areas such as file summarization, code generation, learning assistance, providing inspiration, and even offering life advice and psychological counseling. Building on these observations, we envision a future where social networks are seamlessly integrated with both humans and AIs. We aim to perform community detection within this hybrid network. As far as we are concerned, this concept has not been explored in previous studies, making our work a pioneering effort in this field.

\section{Problem Formulation}
\label{sec:problem formulation}

An HASN graph is denoted as $G(V, E)$, where $\forall v \in V$ is a set of vertices comprising the sets $H$ (human users) and $AI$ (AI entities), such that $|V| = |H| + |AI|$, and $\forall e \in E$ represents the set of edges between humans, AIs, and human-AI connections. 

\textbf{The \problem\ clustering problem} aims to partition an HASN graph into $K$ disjoint subgraphs $C_i(V_i, E_i)$, where $\bigcup_{i=1}^K V_i \subseteq V$ (since AI nodes and their connected edges may be removed during the clustering process) and $V_i \bigcap V_j = \emptyset$, with prior knowledge of which nodes in the network are AI nodes. The goal of \problem\ is to discover a set of clusters (subgraphs) $P = \{ C_i \}_1^K = \{ C_1, C_2, \ldots, C_K \}$ that can maximize human closeness with minimal AI presence. Concretely, a desirable clustering result of an HASN should achieve two key objectives simultaneously: (1) maximizing human closeness and (2) minimizing the presence of AI nodes for each cluster. 

\subsection{Objective Function of \problem}
\label{subsec:objective_function}

To achieve the goal of \problem, we employ a modularity function introduced in a seminal work by Newman as our objective function \cite{newman2004finding}:

\begin{equation}
Q(P=\{C_i\}_{i=1}^K) = \frac{1}{2|E|} \left( \sum_{i=1}^K \sum_{v_p, v_q \in C_i}\left( A_{pq} - \frac{d_p d_q}{2|E|} \right) \right)
\end{equation}
\vspace{0.5em}

Modularity $Q$ measures clustering quality in networks by comparing the density within clusters to the density between clusters. It ranges from -0.5 to 1, with higher scores indicating better clustering. Here, $A$ is the adjacency matrix, $A_{pq}$ indicates the presence of a connection between nodes $p$ and $q$, and $d_p$ is the degree of node $p$. 

To encourage the clustering algorithm to generate cohesive communities with minimal AI presence, we modify the vanilla modularity by infusing a reward-penalty function. This function reweights the clustering quality based on the ratio of humans (and AIs) presence in each cluster $C_i$, defined by:

\begin{equation}
W(C_i) = \beta \cdot \frac{\sum_{v \in C_i} L_v}{|C_i|} - \gamma \cdot \frac{\sum_{v \in C_i} (1 - L_v)}{|C_i|}
\end{equation}
\vspace{0.5em}

\noindent where 
\begin{equation}
L_v =
\begin{cases} 
1, & \text{if node } v \in H \\
0, & \text{if node } v \in AI
\end{cases}
\end{equation}
\vspace{0.5em}

\noindent This leads to a human-centric modularity $HQ$:

\begin{equation}
HQ(P) = \frac{1}{2|E|} \left( \sum_{i=1}^K \alpha \cdot W(C_i) \cdot \left( \sum_{v_p, v_q \in C_i}\left( A_{pq} - \frac{d_p d_q}{2|E|} \right) \right) \right)
\end{equation}
\vspace{0.5em}

\noindent Note that $\beta$ is the weight for rewarding human nodes, $\gamma$ is the weight for penalizing AI nodes, and $\alpha$ is the weight for adjusting the emphasis on human nodes in the objective function \footnote{For simplicity, we set $\alpha$, $\beta$, and $\gamma$ to 1 in our experiments to observe the proposed algorithm’s core behavior without the added complexity of multiple parameters.}. Accordingly, the purpose of \problem\ is to discover a set of clusters (subgraphs) $P = \{ C_i \}_1^K$ that maximizes $HQ$:

\begin{equation}
P^* = \arg \max_{\{C_i\}_{i=1}^k} HQ(\{C_i\}_{i=1}^K)
\end{equation}
\vspace{0.5em}

This objective function promotes the generation of tight-knit communities with minimal AI presence. Since certain AI entities can aid in the formation of these human-centric communities, it is crucial to identify and preserve AI nodes that can promote human closeness while removing those that can not.

\section{Algorithm Design for \problem}
\label{sec:method}

\begin{figure*}[ht]
    \centering
    \includegraphics[height=0.335\linewidth]{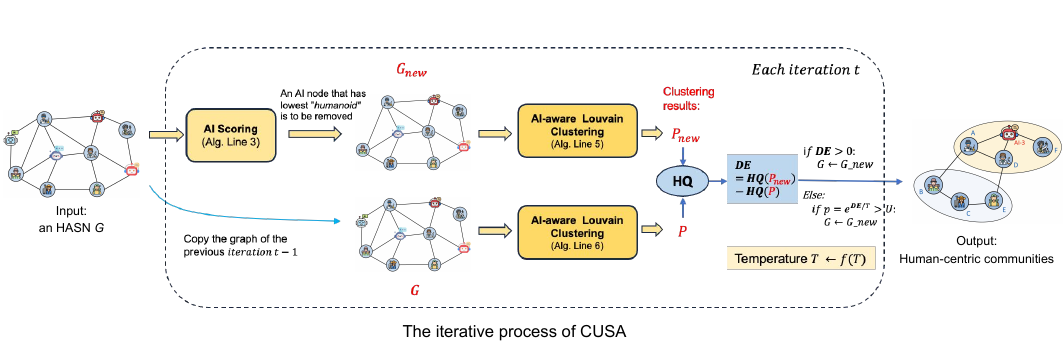}
    \caption{The framework of the proposed CUSA algorithm.}
    \label{figure:CUSA flowchart}
\end{figure*}

To effectively address \problem, we develop a novel algorithm, named \textit{\underline{Cu}stomized AI-aware \underline{S}imulated \underline{A}nnealing} (denoted by CUSA), which integrates three key components (\textit{cf.} Figure \ref{figure:CUSA flowchart}).

\textbf{(i)} To tackle the challenge in the evaluation of AI nodes, CUSA incorporates an \textit{\textbf{AI Scoring}} mechanism to assess the \score\ score of each AI node in a human-AI social network (HASN). CUSA achieves this by assigning distinct weights to edges based on neighboring relationships. Specifically, weights are assigned as follows: human-human edges are denoted by $hh$, human-AI edges by $ha$, and AI-AI edges by $aa$, with the relationship $hh > ha > aa$ (defaults set to 3, 2, and 1, respectively)\footnote{The sensitivity of $hh, ha, aa$ is discussed in Appendix \ref{Sensitivity to hh, ha, aa}.}. CUSA then applies three representative approaches commonly used in social networks—eigenvector centrality, betweenness centrality, and clustering coefficient—to evaluate AI node scores. This encourages the identification and preservation of AI nodes that play significant roles in human-centric communities, aligning with MetaCD’s objectives. Each approach serves a specific purpose:

\noindent (1) Eigenvector Centrality (EC) \cite{freeman2002centrality} considers a node important if it connects to other highly influential nodes. We apply this measure, assuming that AI nodes in the HASN are linked to prominent individuals (or vice versa). Thus, prioritizing AI nodes with high eigenvector centrality in clustering is vital for shaping cohesive community structures and enhancing human closeness.

\noindent (2) Betweenness Centrality (BC) \cite{brandes2001faster} identifies a node as crucial if it serves as a bridge or intermediary within the network. This measure aligns with the objectives of the \problem\ problem, as a beneficial AI node often connects disparate human nodes, potentially reshaping community boundaries and reinforcing human closeness. Therefore, we prioritize preserving AI nodes with high betweenness centrality during clustering \footnote{To mitigate the computational cost of the shortest-path calculations in BC, we utilize an optimized version of the BC evaluation method \cite{brandes2001faster}.}.

\noindent (3) Clustering Coefficient (CC) \cite{watts1998collective}  regards a node as important if its neighbors are densely interconnected. However, in our approach, we prioritize retaining AI nodes with low clustering coefficient scores. This preference is based on the observation that a high clustering coefficient suggests ample interconnectivity within the subnetwork, thereby reducing the necessity for the AI node to act as a bridge among other nodes.

As shown in Table \ref{AI node scoring methods} in the experimental section, the combined use of EC, BC, and CC through a linear combination of their normalized scores yields the most effective results. This indicates that each measure (EC, BC, and CC) captures a distinct aspect of information regarding AI nodes, providing a more comprehensive perspective on their significance within the HASN.

\textbf{(ii)} For the tradeoff between AI removal and community integrity, we develop \textit{\textbf{AI-aware Louvain Clustering Algorithm}} into CUSA. The Louvain clustering algorithm \cite{blondel2008fast} optimizes modularity to detect community structures in networks. Specifically, each node is initially treated as an individual community. For each node $i$ (or community $C_i$), we calculate the modularity gains $\Delta Q$ by moving node $i$ to its neighboring communities $C_j$, is defined as:

\begin{align}
\Delta Q(i \rightarrow C_j) = &\left[ \frac{\Sigma in + k_{i,in}}{2|E|} - \left(\frac{\Sigma tot + k_i}{2|E|}\right)^2 \right] \notag \\
&- \left[ \frac{\Sigma in }{2|E|} - \left(\frac{\Sigma tot}{2|E|}\right)^2 - \left(\frac{k_i}{2|E|}\right)^2 \right]
\end{align}
\vspace{0.5em}

\noindent where $\Sigma in$ is the sum of the weights of the links inside the community $C_j$, $\Sigma tot$ is the sum of the weights of the links incident to nodes in the community $C_j$, $k_i$ is the sum of the weights of the links incident to node $i$, and $k_{i, in}$ is the sum of the weights of the links from node $i$ to nodes in the community $C_j$. At each iteration, communities are aggregated into new super-nodes, condensing into separate communities. This iterative process of evaluating node movements and merging communities continues until no further increase in modularity can be achieved\footnote{The derivation of modularity gain $\Delta Q$ are provided in Appendix \ref{sec: Derivation of delta Q}}.

\begin{algorithm}[t]
\caption{The algorithm of CUSA}
\label{alg:CUSA}
\begin{algorithmic}[1]
\Statex \textbf{Input:}  An HASN $G = (V, E)$ where $|V| = |H| + |AI|$
\Statex \textbf{Output:} $P_{best} = \{ C_i \}_1^K$, the $K$ disjoint clusters (subgraphs) $C_i$, each achieving high human closeness with minimal AI presence (namely "human-centric communities")
\STATE $T \gets T_{initial}$
\WHILE{($T > T_{min}$) \textbf{and} $|AI| > 0$}
    \STATE $rnode \gets \texttt{AIScoring($G$, $AI$)}$
    \STATE $G_{new} \gets \texttt{AIRemove($G, rnode$)}$
    \STATE $P_{new} \gets \texttt{Clustering($G_{new}$)}$
    \STATE $P \gets \texttt{Clustering($G$)}$
    \STATE $DE = HQ(P_{new}) - HQ(P)$
    \IF{$DE > 0$} 
            \STATE $G \gets G_{new}$
    \ELSE 
        \IF{$p = e^{DE / T} > U$ where $U \sim \text{Uniform}(0,1)$}
            \STATE $G \gets G_{new}$ 
        \ENDIF
    \ENDIF
    \IF{$HQ(P) \geq HQ(P_{best})$}
        \STATE $P_{best} \gets P$
        \STATE $T \gets f(T)$
    \ENDIF
\ENDWHILE
\RETURN {} $P_{best}$

\end{algorithmic}
\end{algorithm}

To make the Louvain clustering algorithm more human-specific, we incorporate two reweighting terms into the modularity calculation, $W(C_j)^{before}$ and $W(C_j)^{after}$. These terms (defined in Section 3.1) are based on the human-to-AI ratio within the community $C_j$ before and after merging node $i$ (or community $i$), respectively. This approach makes the AI-aware Louvain clustering algorithm more human-specific by not only considering structural information in each merge step but also emphasizing the proportion of human members within each community. When node $i$ (or community $i$) is merged into community $C_j$, these reweighting terms assess whether the merge increases the proportion of human members, leading to a human-centric modularity gain $\Delta HQ$. This ensures that the resulting communities display a higher human proportion and cohesiveness, fostering a more human-centered clustering outcome. Mathematically, this is achieved with the following formula:

\begin{align}
\Delta HQ(i \rightarrow C_j) = 
&\ \alpha \cdot W(C_j)^{\text{after}} \cdot \left[ \frac{\Sigma \text{in} + k_{i,\text{in}}}{2|E|} 
- \left(\frac{\Sigma \text{tot} + k_i}{2|E|}\right)^2 \right] \nonumber \\
&- \alpha \cdot W(C_j)^{\text{before}} \cdot \left[ \frac{\Sigma \text{in}}{2|E|} 
- \left(\frac{\Sigma \text{tot}}{2|E|}\right)^2 
- \left(\frac{k_i}{2|E|}\right)^2 \right]
\end{align}

\noindent where $\alpha$ is a scaling factor that can adjust the emphasis on the presence of human nodes in the cluster (default set to 1). By iteratively applying this formula, the algorithm effectively identifies a high-modularity partitioning of the network, ensuring that clusters contain a higher proportion of humans.

\textbf{(iii)} To prevent getting trapped in local optima during AI-aware clustering, CUSA infuses the \textit{\textbf{AI-aware Adaptive Clustering (3AC) Framework with a Probability-based Escape Strategy}}. This framework adaptively partitions an HASN by iteratively removing the AI node with the lowest \score, leveraging a probability-based escape strategy to enhance search flexibility. The 3AC framework is central to CUSA’s process, and its steps are as follows:

\begin{enumerate}
    \item Evaluate \score\ for all AI nodes of the HASN graph and rank them.
    \item Apply the probability-based escape strategy to remove the AI node with the lowest \score.
    \item Cluster all remaining nodes to obtain partition $P$.
    \item Calculate the $HQ$ for partition $P$.
    \item Re-evaluate \score\ for all remaining AI nodes and rank them again.
    \item Repeat 2-5 until the $HQ$ of $P$ converges to its highest value.
\end{enumerate}

CUSA precisely implements the core steps of the 3AC framework, advancing through each stage to achieve the framework’s objectives. Specifically, CUSA comprises three major components outlined in the 3AC framework: (1) an evaluation of AI nodes’ importance (referred to as \score), informing the selective retention or removal of AI nodes based on their relevance to human-centric communities; (2) the AI-aware Louvain clustering algorithm, which performs clustering while accounting for the human-to-AI ratio within each community to ensure a human-centric structure; and (3) a probability-based escape strategy that dynamically adjusts the likelihood of removing nodes, helping the algorithm escape local optima and explore the solution space more effectively. Through these integrated components, CUSA iteratively adjusts the community structure by evaluating and balancing the human and AI nodes in each iteration. The pseudo-code for CUSA is presented in Algorithm \ref{alg:CUSA}, while the overall flowchart of the CUSA algorithm is illustrated in Figure \ref{figure:CUSA flowchart} for further understanding. \\

\noindent \textbf{Theorem 4.1.} The time complexity of CUSA is $O(|AI|(|V||E| + |V|^2 \log |V|))$, while the space complexity is $O(|V| + |E|)$.
\begin{proof}
Detailed derivations are provided in Appendix \ref{sec: Computational Complexity of CUSA}.
\end{proof}

\section{Empirical Experiments}
\label{sec:experiments}
\subsection{Experimental Setup}
\label{subsec:experimental_setup}

\noindent\textbf{Benchmark Datasets.} In the absence of real-world HASNs, we propose four strategies for generating such graphs based on the three widely-used datasets (i.e., Cora, CiteSeer, and PubMed). Table \ref{dataset statistics} in Appendix \ref{sec: Basic Statistics of Network Datasets} shows some basic statistics of the three datasets for our experiments.

Initially, we generate a specific number of new AI nodes, such as $n\%$ 
(default set to 10\%) \footnote{The sensitivity to different AI ratios is investigated in Appendix \ref{Sensitivity to Different AI Ratios}.} of the total number of nodes in a real-world network dataset. Subsequently, we employ proposed generation strategies for inserting AI nodes into these networks, thereby constructing HASNs. Notably, after creating the HASNs using the proposed generation strategies, we anticipate the evolution of the social network over time, potentially resulting in the formation of new connections between individuals. Due to space limitations, we outline the four proposed generation strategies along with their specific assumptions \cite{nomiAI2024}\cite{zhang2024better}\cite{skjuve2021my}\cite{loveys2019reducing} and provide detailed descriptions of the evolution method for HASNs in Appendix \ref{sec: The Proposed Four Generation Strategies} and \ref{sec: The Evolution of HASNs}, respectively. Additionally, the validation of generation strategies using real-world HASNs is presented in Appendix \ref{Validation of Generation Strategies}.

\begingroup
\renewcommand{\arraystretch}{1}
\begin{table}[t]
\centering
\caption{The Q, HQ, HMR, and ADM results on the Cora dataset transformed into HASNs using the generation strategy (1), obtained by different AI node scoring methods used in CUSA, compared to no removal and all removal of AIs.}
\LARGE
\resizebox{\columnwidth}{!}{%
\begin{tabular}{lcccc}
\toprule
\textbf{AI Scoring} & \textbf{Q (↑)} & \textbf{HQ (↑)} & \textbf{HMR (↑)} & \textbf{ADM (↑)} \\
\midrule
No removal (N) & 0.7761 & 0.6966 & 42.25\% & 4.24 \\
All removal (A) & 0.8164 & 0.8164 & \underline{0}\% & \underline{0} \\
\midrule
EC & 0.8159 & 0.8092 & 20.34\% & 25.22 \\
BC & 0.8135 & 0.8051 & 27.96\% & 27.32 \\
CC & 0.8156 & 0.8040 & \textbf{58.80}\% & \textbf{43.62} \\
EC+BC & 0.8151 & 0.8069 & 25.63\% & 25.04 \\
EC+CC & 0.8135 & 0.8036 & 31.26\% & 22.58 \\
BC+CC & 0.8113 & 0.8020 & 38.24\% & 27.63 \\
EC+BC+CC & \textbf{0.8190} & \textbf{0.8174} & \underline{14.09}\% & \underline{20.41} \\
\bottomrule
\end{tabular}%
}
\label{AI node scoring methods}
\end{table}
\endgroup

\noindent\textbf{Comparative Methods.} We compare CUSA with four baselines: two traditional non-deep-learning community detection methods (Spectral and Louvain) and two GNN-based community detection methods (GCC and LSEnet).
\begin{enumerate}
    \renewcommand{\labelenumi}{(\theenumi)}
    \item Spectral \cite{amini2013pseudo} is a clustering algorithm based on the normalized Laplacian matrix and the regularized adjacency matrix to minimize ratio cut.
    \item Louvain \cite{blondel2008fast} detects communities by iteratively moving nodes to optimize modularity, maximizing links within communities and minimizing links between them.
    \item GCC \cite{fettal2022efficient} is a graph convolutional clustering method designed for efficient and scalable community detection.
    \item LSEnet \cite{sun2024lsenet} leverages structural embeddings for community detection, integrating node features via manifold-valued graph convolution in hyperbolic space.
    
\end{enumerate}
Each baseline is equipped with three AI node-handling strategies, namely, \underline{n}o removal of AIs (N), \underline{a}ll removal of AIs (A), and removal of certain AIs by GA\underline{D} (D). (\textit{cf.} Figure \ref{illustrative} (a)-(c))
\vspace{1em}

\noindent\textbf{Evaluation Metrics.} For the problem of \problem\ (\textit{cf.} Section \ref{subsec:objective_function}), we employ widely used modularity Q \cite{newman2004finding} and the proposed human-centric modularity HQ as our evaluation metrics to assess the effectiveness of CUSA. The Q accesses clustering quality based on cluster closeness, while HQ does so with minimal AI involvement. Higher Q and HQ (ranging from -0.5 to 1) values indicate better quality of the detected communities. In addition to clustering quality, we aim to understand the impact of remaining AI on community structures after applying CUSA. We propose two metrics: Human Migration Ratio (HMR) and Average AI-driven Migration (ADM). $\text{HMR(\%)} = \frac{\text{\# human node migration}}{\text{\# total human nodes}}$ measures the proportion of humans moving from one community to another, used to evaluate the overall extent of community changes. $\text{ADM} = \frac{\text{\# human node migration}}{\text{\# remaining AI nodes}}$ measures the average number of humans migrating due to remaining AI nodes, used to assess the average influence of AI nodes involved in the community. Higher HMR and ADM values indicate that the preserved AIs have greater capabilities to influence human communities. The next section presents the results, while additional experiments on CUSA scalability, HASNs without evolution, random networks, and real-world HASNs are included in Appendix \ref{sec: Supplementary Experimental Results} for further context.

\begin{table}[t]
\centering
\caption{The HQ results on different datasets each transformed into HASNs using the generation strategy (1), obtained by CUSA in comparison to that of four clustering baselines each equipped with no removal of AIs (N), all removal of AIs (A), and removal of AIs by GAD (D).}
\small
\resizebox{\columnwidth}{!}{%
\begin{tabular}{ccccc}
\toprule
\multicolumn{2}{c}{\textbf{Method}} & \textbf{Cora} & \textbf{CiteSeer} & \textbf{PubMed} \\ 
\hline
\multirow{3}{*}{Spectral \cite{amini2013pseudo}} 
 & N & 0.2907 & 0.2563 & 0.0245\\ 
 & A & 0.3721 & 0.2952 & 0.1394 \\ 
 & D & 0.3308 & 0.2642 & 0.0553 \\ 
 \hline
\multirow{3}{*}{Louvain \cite{blondel2008fast}} 
 & N & 0.6966 & 0.8047 & 0.6613 \\ 
 & A & 0.8164 & 0.8962 & 0.7690 \\  
 & D & 0.7702 & 0.8438 & 0.7245 \\ 
 \hline
\multirow{3}{*}{GCC \cite{fettal2022efficient}} 
 & N & 0.6724 & 0.6353 & 0.6218 \\ 
 & A & 0.7275 & 0.7110 & 0.6672 \\ 
 & D & 0.6910 & 0.6762 & 0.6483 \\ 
 \hline
\multirow{3}{*}{LSEnet \cite{sun2024lsenet}} 
 & N & 0.6628 & 0.6289 & 0.6182 \\ 
 & A & 0.7202 & 0.7089 & 0.6490 \\ 
 & D & 0.6984 & 0.6720 & 0.6378 \\ 
 \hline
\textbf{CUSA} (ours) & - & \textbf{0.8174} & \textbf{0.8984} & \textbf{0.7716} \\ 
\bottomrule
\end{tabular}%
}
\label{The HQ results on different datasets}
\end{table}

\begingroup
\renewcommand{\arraystretch}{1}
\begin{table*}[t]
\centering
\caption{The Q and HQ results on the Cora dataset transformed into HASNs using different generation strategies, obtained by CUSA in comparison to that of four clustering baselines each equipped with no removal of AIs (N), all removal of AIs (A), and removal of AIs by GAD (D).}
\label{The Q and HQ results on the Cora dataset}
\resizebox{\textwidth}{!}{%
\begin{tabular}{cccccccccc}
\toprule
\multirow{2}{*}{\makebox[0.15\textwidth]{\textbf{Method}}} & \multirow{2}{*}{} & \multicolumn{2}{c}{\textbf{\textit{k\%} random insertion}} & \multicolumn{2}{c}{\textbf{Inverse degree insertion}} & \multicolumn{2}{c}{\textbf{Inner and outer AI mix}} & \multicolumn{2}{c}{\textbf{AI with dual personality}} \\
\cline{3-10}
& & \textbf{Q (↑)} & \textbf{HQ (↑)} & \textbf{Q (↑)} & \textbf{HQ (↑)} & \textbf{Q (↑)} & \textbf{HQ (↑)} & \textbf{Q (↑)} & \textbf{HQ (↑)} \\
\hline
\multirow{3}{*}{\makebox[0.15\textwidth]{Spectral \cite{amini2013pseudo}}} 
& N & 0.3087 & 0.2907 & 0.3878 & 0.3733 & 0.3013 & 0.2833 & 0.3206 & 0.3028 \\
& A & 0.3721 & 0.3721 & 0.3958 & 0.3958 & 0.3553 & 0.3552 & 0.3594 & 0.3594 \\
& D & 0.3378 & 0.3308 & 0.3902 & 0.3818 & 0.3148 & 0.3066 & 0.3334 & 0.3240 \\
\hline
\multirow{3}{*}{\makebox[0.15\textwidth]{Louvain \cite{blondel2008fast}}} 
& N & 0.7761  & 0.6866 & 0.8123 & 0.7561 & 0.8265 & 0.7994 & 0.8094 & 0.7688 \\
& A & 0.8164 & 0.8164 & 0.8235 & 0.8235 & 0.8214 & 0.8214 & 0.8202 & 0.8202 \\
& D & 0.7990 & 0.7702 & 0.8182 & 0.7895 & 0.8238 & 0.8070 & 0.8152 & 0.7938 \\
\hline
\multirow{3}{*}{\makebox[0.15\textwidth]{GCC \cite{fettal2022efficient}}} 
& N & 0.6612 & 0.6243 & 0.7032 &0.6795 & 0.7186 & 0.6824 & 0.7218 & 0.6920 \\
& A & 0.7275 & 0.7275 & 0.7462 & 0.7462 & 0.7523 & 0.7523 & 0.7508 & 0.7508 \\
& D & 0.6846 & 0.6731 & 0.7315 & 0.7168 & 0.7420 & 0.7325 & 0.7436 & 0.7296 \\
\hline
\multirow{3}{*}{\makebox[0.15\textwidth]{LSEnet \cite{sun2024lsenet}}} 
& N & 0.6973 & 0.6342 & 0.7284 & 0.6857 & 0.7514 & 0.7269 & 0.7213 & 0.7029 \\
& A & 0.7202 & 0.7202 & 0.7423  & 0.7423 & 0.7558  & 0.7558 & 0.7412 & 0.7412 \\
& D & 0.7023 & 0.6876 & 0.7370 & 0.7152 & 0.7520 & 0.7325 & 0.7364 & 0.7283 \\
\hline
\multirow{1}{*}{\makebox[0.15\textwidth]{\textbf{CUSA} (ours)}} 
& - & \textbf{0.8190} & \textbf{0.8174} & \textbf{0.8251} & \textbf{0.8245} & \textbf{0.8310} & \textbf{0.8224} & \textbf{0.8214} & \textbf{0.8211} \\
\bottomrule
\end{tabular}%
}
\end{table*}
\endgroup

\subsection{Experimental Results}
\subsubsection{Comparison of Different AI Node Scoring Methods Used in CUSA} In the first set of experiments, we assess the performance of different AI scoring methods used in CUSA compared to standard Louvain clustering \cite{blondel2008fast} without AI removal and with all AI removal, when clustering on the Cora dataset transformed into HASNs using the generation strategy (1). We test the three different scoring methods, eigenvector centrality (EC), betweenness centrality (BC), and clustering coefficient (CC), as well as their combination scores (denoted by "+" for the linear combination of normalized scores). The corresponding results are presented in Table \ref{AI node scoring methods} \footnote{The HMR and ADM results on different datasets are examined in Appendix \ref{HMR and ADM Results on Different Datasets}.}. Inspection of Table \ref{AI node scoring methods} reveals three noteworthy points. First, all seven AI scoring methods (below the third line) used in CUSA outperform the standard Louvain clustering with no AI removal, as shown by higher Q and HQ scores, demonstrating their better ability to capture structural nuances in HASNs. Second, the EC+BC+CC stands out in performance, achieving a 5.3\% and 17.6\% relative gain in Q and HQ, respectively, compared to clustering with no AI removal. Moreover, the HMR and ADM of the baseline method with all AI removal are 0\% and 0, respectively, whereas our CUSA achieves 14.09\% and 20.41. These suggest that EC, BC, and CC capture different aspects of information about AI nodes. Third, CC excels in HMR and ADM, indicating that CC may serve as a bridge for potential communities. In the following experiments, we adopt EC+BC+CC which achieves the best Q and HQ scores, as our AI scoring in CUSA (i.e., line 3 of Algorithm \ref{alg:CUSA}) for comparison with the baselines \footnote{Due to the randomness in HASNs generated by our strategies, each experiment was run 100 times, and the results were averaged for reliability.}.

\subsubsection{Comparison of CUSA with Baselines on Different Datasets} In the second set of experiments, we compare CUSA against four baselines: two traditional non-deep learning methods, Spectral \cite{amini2013pseudo} and Louvain \cite{blondel2008fast}, and two GNN-based methods, GCC \cite{fettal2022efficient} and LSEnet \cite{sun2024lsenet}. Evaluations are conducted on three datasets (Cora, CiteSeer, and PubMed), each transformed into HASNs using the proposed generation strategy (1). Notably, we equip each baseline with three AI node-handling strategies: \underline{n}o removal of AIs (N), \underline{a}ll removal of AIs (A), and removal of certain AIs by GA\underline{D} techniques (D) (\textit{cf.} Figure \ref{illustrative}. (a)-(c)). As to the GAD method, we adopt the recently proposed DCI method \cite{wang2021decoupling}, which leverages self-supervised learning to decouple node representation learning from classification, to detect anomalous AI nodes in an HASN. Two important observations can be drawn from Table \ref{The HQ results on different datasets} which presents the human-centric HQ scores (modified from standard modularity Q in Section \ref{subsec:objective_function}) of the resulting clusters.
First, across the three datasets, each baseline achieves the highest HQ when all AI nodes are removed, followed by GAD-based removal, while the worst occurs with no AI removal. Second, CUSA significantly outperforms baselines across three datasets by adaptively balancing AI removal and community integrity, leading to better clustering results. In contrast, baselines struggle to handle the interweaving of humans and AIs.

\subsubsection{Comparison of CUSA with Baselines on a Dataset using Different Generation Strategies} To further validate the superiority and feasibility of CUSA, we evaluate its performance on the Cora dataset, transformed into four HASNs using the proposed generation strategies: $k\%$ random insertion, inverse degree insertion, inner and outer AI mix, and AI with dual personality (\textit{cf.} Appendix \ref{sec: The Proposed HASN Scenarios}). Each AI-augmented Cora graph evolves through link prediction based on Jaccard similarity \cite{arrar2024comprehensive}\cite{daud2020applications}. We compare CUSA against four baselines, each equipped with three AI node-handling strategies: \underline{n}o removal of AIs (N), \underline{a}ll removal of AIs (A), and removal of certain AIs by GA\underline{D} (D). We present the two evaluation metrics standard Q and human-centric HQ (\textit{cf.} Section \ref{subsec:objective_function}). The corresponding results are shown in Table \ref{The Q and HQ results on the Cora dataset}, from which we can draw three important observations. First, across four HASN scenarios, each baseline obtains the highest Q and HQ when all AI nodes are removed, followed by GAD-based removal, with the worst performance when no AI nodes are removed. Second, CUSA outperforms all baselines, achieving an average relative gain of 21.8\% in Q and HQ (e.g., inverse degree insertion with full AI removal). These indeed demonstrate the efficacy of AI-aware clustering techniques for dealing with \problem\ problem since they identify and preserve AI nodes that can potentially reshape new communities and enhance human closeness. Third, as we can see, the elaborate generation strategies of the latter three result in clusters with significantly better performance compared to random insertion ($k\%$ random insertion). This also suggests that industries generating AIs can use these strategies to enhance human closeness and discover potential communities for such emerging social networks (HASNs) in the future.

\section{Conclusion}
\label{sec:conclusion}
To the best of our knowledge, this paper is the first to investigate human-centric community detection in hybrid networks (HASNs) that contain human and AI nodes. Based on four proposed HASN scenarios, we introduce a new problem, \problem, which aims to identify clusters that maximize human closeness while minimizing AI presence. To address \problem, we develop a novel algorithm, CUSA, which leverages AI-aware clustering techniques to balance the trade-off between AI removal and community integrity. Empirical evaluations on real social networks, reconfigured as HASNs, demonstrate the effectiveness of CUSA compared with state-of-the-art methods. Furthermore, we observe that tailored generation strategies can further enhance clustering outcomes, providing valuable insights for enterprises developing AIs to foster human connections and uncover latent communities.

\section*{Acknowledgments}
This work is supported in part by the NSTC, Taiwan, under grants 113-2223-E-002-011, 113-2221-E-001-016-MY3, 112-2221-E-001-010-MY3, and 111-2221-E-002-135-MY3, by the Ministry of Education, Taiwan, under grant MOE 113L9009, and by Academia Sinica, Taiwan, under Academia Sinica Investigator Project Grant AS-IV-114-M06. We thank the NCHC for providing computational and storage resources.

\bibliographystyle{unsrt}
\bibliography{reference}

\begin{thebibliography}{10}

\bibitem{mcpherson2001birds}
Miller McPherson, Lynn Smith-Lovin, and James~M Cook.
\newblock Birds of a feather: Homophily in social networks.
\newblock {\em Annual review of sociology}, 27(1):415--444, 2001.

\bibitem{su2022comprehensive}
Xing Su, Shan Xue, Fanzhen Liu, Jia Wu, Jian Yang, Chuan Zhou, Wenbin Hu, Cecile Paris, Surya Nepal, Di~Jin, et~al.
\newblock A comprehensive survey on community detection with deep learning.
\newblock {\em IEEE Transactions on Neural Networks and Learning Systems}, 2022.

\bibitem{jin2021survey}
Di~Jin, Zhizhi Yu, Pengfei Jiao, Shirui Pan, Dongxiao He, Jia Wu, S~Yu Philip, and Weixiong Zhang.
\newblock A survey of community detection approaches: From statistical modeling to deep learning.
\newblock {\em IEEE Transactions on Knowledge and Data Engineering}, 35(2):1149--1170, 2021.

\bibitem{du2007community}
Nan Du, Bin Wu, Xin Pei, Bai Wang, and Liutong Xu.
\newblock Community detection in large-scale social networks.
\newblock In {\em Proceedings of the 9th WebKDD and 1st SNA-KDD 2007 workshop on Web mining and social network analysis}, pages 16--25, 2007.

\bibitem{umrawal2023community}
Abhishek~K Umrawal, Christopher~J Quinn, and Vaneet Aggarwal.
\newblock A community-aware framework for social influence maximization.
\newblock {\em IEEE Transactions on Emerging Topics in Computational Intelligence}, 7(4):1253--1262, 2023.

\bibitem{meijers2021globalxr}
A.~Meijers.
\newblock {Global XR Conference: The Largest Global XR Community Event in the World}.
\newblock \url{https://techcommunity.microsoft.com/t5/mixed-reality-blog/global-xr-conference-the-largest-global-xr-community-event-in/ba-p/3043539}, 2021.
\newblock Accessed: 2022-12-15.

\bibitem{meta2023workrooms}
{Meta}.
\newblock Workrooms.
\newblock \url{https://www.oculus.com/workrooms/?utm_medium=referral&utm_source=xr4work.com}, 2023.
\newblock Accessed: 2021-11-03.

\bibitem{nomiAI2024}
Nomi AI.
\newblock I have a group chat with three {AI} friends, thanks to nomi ai — they’re getting too smart.
\newblock https://reurl.cc/MOeN3m, 2024.
\newblock Accessed: 2024-07-06.

\bibitem{engage2023}
{ENGAGE}.
\newblock {ENGAGE}.
\newblock \url{https://engagevr.io/}, 2023.
\newblock Accessed: 2023-01-15.

\bibitem{hubs2023mozilla}
{Hubs}.
\newblock Hubs mozilla.
\newblock \url{https://hubs.mozilla.com/?utm_medium=referral&utm_source=xr4work.com}, 2023.
\newblock Accessed: 2021-11-03.

\bibitem{ma2021comprehensive}
Xiaoxiao Ma, Jia Wu, Shan Xue, Jian Yang, Chuan Zhou, Quan~Z Sheng, Hui Xiong, and Leman Akoglu.
\newblock A comprehensive survey on graph anomaly detection with deep learning.
\newblock {\em IEEE Transactions on Knowledge and Data Engineering}, 35(12):12012--12038, 2021.

\bibitem{lamichhane2024anomaly}
Prabin~B Lamichhane and William Eberle.
\newblock Anomaly detection in graph structured data: A survey.
\newblock {\em arXiv preprint arXiv:2405.06172}, 2024.

\bibitem{techcrunch_meta_2024}
{TechCrunch}.
\newblock Meta plans to bring generative {AI} to metaverse games.
\newblock https://reurl.cc/r9yAXZ, jul 2024.
\newblock Accessed: 2024-07-10.

\bibitem{wang2024survey}
Lei Wang, Chen Ma, Xueyang Feng, Zeyu Zhang, Hao Yang, Jingsen Zhang, Zhiyuan Chen, Jiakai Tang, Xu~Chen, Yankai Lin, et~al.
\newblock A survey on large language model based autonomous agents.
\newblock {\em Frontiers of Computer Science}, 18(6):186345, 2024.

\bibitem{amini2013pseudo}
Arash~A Amini, Aiyou Chen, Peter~J Bickel, and Elizaveta Levina.
\newblock Pseudo-likelihood methods for community detection in large sparse networks.
\newblock 2013.

\bibitem{blondel2008fast}
Vincent~D Blondel, Jean-Loup Guillaume, Renaud Lambiotte, and Etienne Lefebvre.
\newblock Fast unfolding of communities in large networks.
\newblock {\em Journal of statistical mechanics: theory and experiment}, 2008(10):P10008, 2008.

\bibitem{kirkpatrick1983optimization}
Scott Kirkpatrick, C~Daniel Gelatt~Jr, and Mario~P Vecchi.
\newblock Optimization by simulated annealing.
\newblock {\em science}, 220(4598):671--680, 1983.

\bibitem{boettcher2002optimization}
Stefan Boettcher and Allon~G Percus.
\newblock Optimization with extremal dynamics.
\newblock {\em complexity}, 8(2):57--62, 2002.

\bibitem{newman2013spectral}
Mark~EJ Newman.
\newblock Spectral methods for community detection and graph partitioning.
\newblock {\em Physical Review E—Statistical, Nonlinear, and Soft Matter Physics}, 88(4):042822, 2013.

\bibitem{lecun2015deep}
Yann LeCun, Yoshua Bengio, and Geoffrey Hinton.
\newblock Deep learning.
\newblock {\em nature}, 521(7553):436--444, 2015.

\bibitem{khoshraftar2024survey}
Shima Khoshraftar and Aijun An.
\newblock A survey on graph representation learning methods.
\newblock {\em ACM Transactions on Intelligent Systems and Technology}, 15(1):1--55, 2024.

\bibitem{tsitsulin2023graph}
Anton Tsitsulin, John Palowitch, Bryan Perozzi, and Emmanuel M{\"u}ller.
\newblock Graph clustering with graph neural networks.
\newblock {\em Journal of Machine Learning Research}, 24(127):1--21, 2023.

\bibitem{ge2024unsupervised}
Youming Ge, Cong Huang, Yubao Liu, Sen Zhang, and Weiyang Kong.
\newblock Unsupervised social network embedding via adaptive specific mappings.
\newblock {\em Frontiers of Computer Science}, 18(3):183310, 2024.

\bibitem{chikwendu2023comprehensive}
Ijeoma~Amuche Chikwendu et~al.
\newblock A comprehensive survey on deep graph representation learning methods.
\newblock {\em Journal of Artificial Intelligence Research}, 78:287--356, 2023.

\bibitem{wang2023overview}
Shiping Wang, Jinbin Yang, Jie Yao, Yang Bai, and William Zhu.
\newblock An overview of advanced deep graph node clustering.
\newblock {\em IEEE Transactions on Computational Social Systems}, 11(1):1302--1314, 2023.

\bibitem{pan2021multi}
Erlin Pan and Zhao Kang.
\newblock Multi-view contrastive graph clustering.
\newblock {\em Advances in neural information processing systems}, 34:2148--2159, 2021.

\bibitem{fang2020survey}
Yixiang Fang, Xin Huang, Lu~Qin, Ying Zhang, Wenjie Zhang, Reynold Cheng, and Xuemin Lin.
\newblock A survey of community search over big graphs.
\newblock {\em The VLDB Journal}, 29:353--392, 2020.

\bibitem{wu2020deep}
Ling Wu, Qishan Zhang, Chi-Hua Chen, Kun Guo, and Deqin Wang.
\newblock Deep learning techniques for community detection in social networks.
\newblock {\em IEEE Access}, 8:96016--96026, 2020.

\bibitem{shao2024distributed}
Yingxia Shao, Hongzheng Li, Xizhi Gu, Hongbo Yin, Yawen Li, Xupeng Miao, Wentao Zhang, Bin Cui, and Lei Chen.
\newblock Distributed graph neural network training: A survey.
\newblock {\em ACM Computing Surveys}, 56(8):1--39, 2024.

\bibitem{pan2024identification}
Yu~Pan, Feng Yao, Xin Liu, Lei Zhang, Wei Li, and Pei Wang.
\newblock Identification of dynamic networks community by fusing deep learning and evolutionary clustering.
\newblock 2024.

\bibitem{bilot2023graph}
Tristan Bilot, Nour El~Madhoun, Khaldoun Al~Agha, and Anis Zouaoui.
\newblock Graph neural networks for intrusion detection: A survey.
\newblock {\em IEEE Access}, 11:49114--49139, 2023.

\bibitem{xiang2023semi}
Sheng Xiang et~al.
\newblock Semi-supervised credit card fraud detection via attribute-driven graph representation.
\newblock In {\em Proceedings of the AAAI Conference on Artificial Intelligence}, volume~37, pages 14557--14565, 2023.

\bibitem{yu2016survey}
Rose Yu, Huida Qiu, Zhen Wen, ChingYung Lin, and Yan Liu.
\newblock A survey on social media anomaly detection.
\newblock {\em ACM SIGKDD Explorations Newsletter}, 18(1):1--14, 2016.

\bibitem{roy2024gad}
Amit Roy, Juan Shu, Jia Li, Carl Yang, Olivier Elshocht, Jeroen Smeets, and Pan Li.
\newblock Gad-nr: Graph anomaly detection via neighborhood reconstruction.
\newblock In {\em Proceedings of the 17th ACM International Conference on Web Search and Data Mining}, pages 576--585, 2024.

\bibitem{guo2024large}
Taicheng Guo, Xiuying Chen, Yaqi Wang, Ruidi Chang, Shichao Pei, Nitesh~V Chawla, Olaf Wiest, and Xiangliang Zhang.
\newblock Large language model based multi-agents: A survey of progress and challenges.
\newblock {\em arXiv preprint arXiv:2402.01680}, 2024.

\bibitem{achiam2023gpt}
Josh Achiam et~al.
\newblock Gpt-4 technical report.
\newblock {\em arXiv preprint arXiv:2303.08774}, 2023.

\bibitem{team2023gemini}
Gemini Team et~al.
\newblock Gemini: a family of highly capable multimodal models.
\newblock {\em arXiv preprint arXiv:2312.11805}, 2023.

\bibitem{AnthropicAI2023}
AnthropicAI.
\newblock Introducing claude.
\newblock 2023.
\newblock Retrieved from https://www.anthropic.com/index/introducing-claude.

\bibitem{newman2004finding}
Mark~EJ Newman and Michelle Girvan.
\newblock Finding and evaluating community structure in networks.
\newblock {\em Physical review E}, 69(2):026113, 2004.

\bibitem{freeman2002centrality}
Linton~C Freeman et~al.
\newblock Centrality in social networks: Conceptual clarification.
\newblock {\em Social network: critical concepts in sociology. Londres: Routledge}, 1:238--263, 2002.

\bibitem{brandes2001faster}
Ulrik Brandes.
\newblock A faster algorithm for betweenness centrality.
\newblock {\em Journal of mathematical sociology}, 25(2):163--177, 2001.

\bibitem{watts1998collective}
Duncan~J Watts and Steven~H Strogatz.
\newblock Collective dynamics of ‘small-world’networks.
\newblock {\em nature}, 393(6684):440--442, 1998.

\bibitem{zhang2024better}
Jie Zhang, Dongrui Liu, Chen Qian, Ziyue Gan, Yong Liu, Yu~Qiao, and Jing Shao.
\newblock The better angels of machine personality: How personality relates to llm safety.
\newblock {\em arXiv preprint arXiv:2407.12344}, 2024.

\bibitem{skjuve2021my}
Marita Skjuve, Asbj{\o}rn F{\o}lstad, Knut~Inge Fostervold, and Petter~Bae Brandtzaeg.
\newblock My chatbot companion-a study of human-chatbot relationships.
\newblock {\em International Journal of Human-Computer Studies}, 149:102601, 2021.

\bibitem{loveys2019reducing}
Kate Loveys et~al.
\newblock Reducing patient loneliness with artificial agents: design insights from evolutionary neuropsychiatry.
\newblock {\em Journal of medical Internet research}, 21(7):e13664, 2019.

\bibitem{fettal2022efficient}
Chakib Fettal, Lazhar Labiod, and Mohamed Nadif.
\newblock Efficient graph convolution for joint node representation learning and clustering.
\newblock {\em Proceedings of the fifteenth ACM International conference on web search and data mining (WSDM)}, pages 289--297, 2022.

\bibitem{sun2024lsenet}
Li~Sun, Zhenhao Huang, Hao Peng, Yujie Wang, Chunyang Liu, and Philip~S Yu.
\newblock Lsenet: Lorentz structural entropy neural network for deep graph clustering.
\newblock {\em Proceedings of the 41 st International Conference on Machine Learning (ICML)}, 2024.

\bibitem{wang2021decoupling}
Yanling Wang et~al.
\newblock Decoupling representation learning and classification for gnn-based anomaly detection.
\newblock In {\em Proceedings of the 44th international ACM SIGIR conference on research and development in information retrieval}, pages 1239--1248, 2021.

\bibitem{arrar2024comprehensive}
Djihad Arrar et~al.
\newblock A comprehensive survey of link prediction methods.
\newblock 80(3):3902--3942, 2024.

\bibitem{daud2020applications}
Nur~Nasuha Daud, Siti~Hafizah Ab~Hamid, Muntadher Saadoon, Firdaus Sahran, and Nor~Badrul Anuar.
\newblock Applications of link prediction in social networks: A review.
\newblock {\em Journal of Network and Computer Applications}, 166:102716, 2020.

\bibitem{sen2008collective}
Prithviraj Sen et~al.
\newblock Collective classification in network data.
\newblock {\em AI magazine}, 29(3):93--93, 2008.

\bibitem{zhang2022new}
Min Zhang, Xiaojuan Wang, Lei Jin, Mei Song, and Ziyang Li.
\newblock A new approach for evaluating node importance in complex networks via deep learning methods.
\newblock {\em Neurocomputing}, 497:13--27, 2022.

\bibitem{de2021centrality}
Omar De~la Cruz~Cabrera, Mona Matar, and Lothar Reichel.
\newblock Centrality measures for node-weighted networks via line graphs and the matrix exponential.
\newblock {\em Numerical Algorithms}, 88:583--614, 2021.

\bibitem{yang2012defining}
Jaewon Yang and Jure Leskovec.
\newblock Defining and evaluating network communities based on ground-truth.
\newblock In {\em Proceedings of the ACM SIGKDD workshop on mining data semantics}, pages 1--8, 2012.

\bibitem{erdds1959random}
P~ERDdS and A~R\&wi.
\newblock On random graphs i.
\newblock {\em Publ. math. debrecen}, 6(290-297):18, 1959.

\bibitem{chen2022user}
Bing-Jyue Chen and De-Nian Yang.
\newblock User recommendation in social metaverse with vr.
\newblock In {\em Proceedings of the 31st ACM International Conference on Information \& Knowledge Management}, pages 148--158, 2022.

\end{thebibliography}
\clearpage

\appendix

\begin{figure*}[t]
    \centering
    \includegraphics[height=0.228\linewidth]{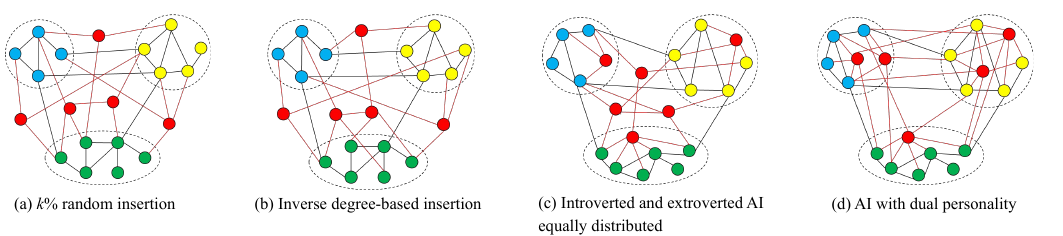}
    \caption{The four proposed hybrid network (HASN) scenarios, where red circles represent AI nodes and other colors represent human nodes.}
    \label{figure:4_scenario}
\end{figure*}

\section*{Appendix}

\section{Potential Application Scenarios}
\label{sec: Potential Application Scenarios}

We discuss several potential real-world applications for MetaCD, showcasing how it enhances user experiences across entertainment, psychology, and e-learning. By complementing human interactions with AI nodes, MetaCD fosters collaboration, builds connections, and supports diverse needs while addressing challenges to ensure effective integration.

\noindent \textbf{(1) Entertainment:} \textit{Enhancing Social Metaverse Platforms}
\\Social metaverse platforms (e.g., Roblox and IMVU) can leverage MetaCD to help users discover suitable communities. AI nodes, tailored to individual preferences, seamlessly participate in social activities and fill gaps left by unavailable human participants, thereby enhancing the overall user experience.

For instance, Roblox has introduced AI companions that facilitate large-scale social interactions and actively engage in group activities. These companions assist humans in completing collaborative quests within the metaverse. When two previously unacquainted individuals repeatedly collaborate with the same AI companion during group tasks, they are more likely to interact and form connections to complete the task more effectively. Furthermore, AI companions can help maintain group functionality even when there are insufficient human members. However, including too many AI companions in a group can negatively impact the user experience. Group quests in the metaverse are designed to help humans develop skills and foster collaboration, not to rely solely on AI companions. An excessive number of AI participants may overshadow human contributions, resulting in fewer rewards for humans and discouraging their participation in future quests. Balancing the number of AI companions to facilitate interaction without overshadowing human contributions is therefore crucial for maintaining a positive user experience and encouraging continued engagement in the social metaverse.

\noindent \textbf{(2) Psychology:} \textit{Building Supportive Online Therapy Groups}\\ 
Online therapy services (e.g., 7 Cups and ReachOut) can utilize MetaCD to form effective support groups. AI nodes serve as empathetic companions, fostering open, warm, and judgment-free discussions to promote emotional well-being.

For example, 7 Cups integrates AI companions trained with diverse personalities—some calm and rational, others passionate and emotional—making them suitable for engaging with different types of participants. These AI companions observe participants’ reactions, exchange insights with other AI companions, and determine the best ways to guide individuals. They encourage participants to share their thoughts and care for one another, helping to build meaningful social connections. However, if a support group includes too many AI participants, the opportunities for human-to-human experience-sharing diminish. This can lead to a sense of artificiality and erode confidence in the coping strategies learned during group interactions, as participants may doubt their real-world applicability. To ensure the effectiveness of the support group, it is essential to maintain an appropriate balance of AI companions. This balance enables participants to receive emotional support and practical coping strategies while fostering genuine social relationships.

\noindent \textbf{(3) E-Learning:} \textit{Facilitating Collaborative Study Groups}\\
E-learning platforms (e.g., Khan Academy and Duolingo) can employ MetaCD to help students form effective study groups. AI nodes act as tireless teaching assistants, providing continuous guidance and engaging as peers at similar skill levels, fostering mutual motivation and sustained learning.

For instance, Khan Academy integrates AI companions that eagerly answer questions, guide problem-solving, and create tailored study plans. These AI companions collaborate within the study group to track learners’ progress and ensure that the group advances steadily. They encourage learners to engage in discussions, assist each other in solving problems, and identify potential long-term learning partners. Despite their benefits, an overabundance of AI companions in a study group can reduce opportunities for learners to gain insights from peers with relevant experiences, such as balancing different subjects or managing study time effectively. This lack of shared experiences can make the group dynamic feel monotonous, leading to reduced engagement. Therefore, maintaining an optimal number of AI companions is key to fostering mutual learning and motivation while ensuring that human learners actively contribute to and benefit from the study group.

\section{The Derivation of $\Delta Q$}
\label{sec: Derivation of delta Q}

We provide a combinatorial interpretation for the derivation of modularity gain $\Delta Q$ as follows. First, we reformulate the modularity $Q$ function:
\begin{align}
Q(P = C_i^K) = 
&\ \frac{1}{2|E|} \left( \sum_{i=1}^{K} \sum_{v_p, v_q \in C_i} \left( A_{pq} - \frac{d_p d_q}{2|E|} \right) \right) \nonumber \\
= 
&\ \frac{1}{2|E|} \sum_{i=1}^{K} \left( \frac{\Sigma_{\text{in}}^i}{2|E|} - \left( \frac{\Sigma_{\text{tot}}^i}{2|E|} \right)^2 \right),
\end{align}

\noindent where $\Sigma_{\text{in}}^i$ is the sum of the weights of the links inside the community $C_i$, and $\Sigma_{\text{tot}}^i$ is the sum of the weights of the links incident to nodes in $C_i$.

Note that the modularity gain $\Delta Q$ is obtained by moving the node $i$ (which is treated as an individual community $C_i$) into the community $C_j$. By the reformulated modularity function $Q$, to obtain the modularity gains $\Delta Q$, we only need to consider the communities $C_i$ and $C_j$. Since the node $i$ is moved to the community $C_j$, the sum of the weights of the links inside the community $C_j$, i.e., $\Sigma_{\text{in}}^j$, is updated to $\Sigma_{\text{in}}^j + k_{i, \text{in}}^j$, where $k_{i, \text{in}}^j$ is the sum of the weights of the links from node $i$ to nodes in the community $C_j$. Likewise, the sum of the weights of the links incident to nodes in the community $C_j$, i.e., $\Sigma_{\text{tot}}^j$, is updated to $\Sigma_{\text{tot}}^j + k_i$, where $k_i$ is the sum of the weights of the links incident to node $i$.

On the other hand, since the node $i$ is treated as an individual community $C_i$ and is moved to the community $C_j$, the modularity gain is $\left( \frac{k_i}{2|E|} \right)^2$. Therefore, the modularity gain $\Delta Q$ is:
\begin{align}
\Delta Q = 
\frac{\Sigma_{\text{in}}^j + k_{i, \text{in}}^j}{2|E|} 
- \left( \frac{\Sigma_{\text{tot}}^j + k_i}{2|E|} \right)^2 - \frac{\Sigma_{\text{in}}^j}{2|E|} 
+ \left( \frac{\Sigma_{\text{tot}}^j}{2|E|} \right)^2 
+ \left( \frac{k_i}{2|E|} \right)^2.
\end{align}

\section{Computational Complexity of CUSA}
\label{sec: Computational Complexity of CUSA}

\noindent \textbf{Time Complexity.} For each iteration, CUSA first evaluates AI node scores in $O(|V||E|) + |V|^2 \log |V|$ time on weighted graphs, where the eigenvector centrality is known to have complexity $O(|V|^2)$. The betweenness centrality is obtained in $O(|V||E|) + |V|^2 \log |V|$ time by computing the shortest paths for all pairs (augmented Dijkstra), and the clustering coefficient is evaluated in $O(|V| \cdot d_{\text{max}}^2)$ time by adjacency list intersection ($d_{\text{max}}$ is the size of the largest adjacency list of all vertices in the graph). Then, CUSA clusters the graph in $O(|V| \log |V|)$ time by the Louvain Clustering Algorithm and computes HQ in $O(|V|^2)$ time by considering all $(v_p, v_q)$ pairs. Since an AI node must be removed in each iteration, there are at most $|AI|$ iterations. Therefore, the time complexity is $O(|AI|(|V||E| + |V|^2 \log |V|))$.

\noindent \textbf{Space Complexity.} CUSA evaluates AI node scores and clusters the graph with $O(|V| + |E|)$ space (using the adjacency list to represent the graph data structure). Therefore, the space complexity is $O(|V| + |E|)$.

\section{The Proposed HASN Scenarios}
\label{sec: The Proposed HASN Scenarios}

\subsection{The Proposed Four Generation Strategies}
\label{sec: The Proposed Four Generation Strategies}

We outline the four proposed generation strategies for synthesizing HASNs with the specific assumptions  \cite{nomiAI2024}\cite{zhang2024better}\cite{skjuve2021my}\cite{loveys2019reducing}, as depicted in Figure \ref{figure:4_scenario}. These strategies reflect various interaction patterns between human and AI nodes, laying the foundation for constructing meaningful human-AI social networks (HASNs) for our experiments. 

\noindent \textbf{(1) \textit{k\% random insertion}}: AI interacts randomly with humans (or other AIs) with a uniform distribution of $k\%$. This setup assumes that everyone in the evolving \XR\ has equal chances to interact with AI, as depicted in Figure \ref{figure:4_scenario} (a).

\noindent \textbf{(2) \textit{Inverse degree-based insertion}}: The probability of an AI interacting with humans (or other AIs) is inversely proportional to the node's degree. This can be modeled using an exponential decay function, given by ${prob\_connect}(v_i) = a \times e^{-r \times (d_{v_i} - 1)}$, where $d_{v_i}$ is the degree of node $v_i$, $a$ is the initial probability (i.e., when $d_{v_i}=1$), and $r$ is the decay constant. This indicates that nodes with lower degrees are more likely to be linked by AI. This setup assumes that certain individuals tend toward introversion and prefer interactions with AI virtual entities, as depicted in Figure \ref{figure:4_scenario} (b).

\noindent \textbf{(3) \textit{Introverted and extroverted AI equally distributed}}: This suggests the existence of two distinct AI types: introverted and extroverted. Introverted AI engages extensively with members in their respective groups with a probability of $x\%$, while extroverted ones interact with members across various groups with a probability of $y\%$ (typically with $x>y$). This setup assumes that half of the AIs engage primarily with group members due to shared interests, while the other half possess versatility, enabling interaction across multiple groups, as depicted in Figure \ref{figure:4_scenario} (c).

\noindent \textbf{(4) \textit{AI with dual personality}}: This suggests that AI may demonstrate dual personality, actively engaging with members within its own community with a probability of $x\%$, while also interacting with individuals in other communities with a probability of $y\%$ (typically with $x>y$). This setup assumes that AIs adeptly interact with community members due to shared interests and are also versatile enough to engage with individuals from other communities, as depicted in Figure \ref{figure:4_scenario} (d).

\subsection{The Evolution of HASNs}
\label{sec: The Evolution of HASNs}
To better simulate real-world behavior, AI nodes in social networks are designed to take time to generate content, establish links, and engage with communities. This leads to the formation of a mixed human-AI social network (HASN) rather than simply inserting AI nodes into an existing network. AI nodes are introduced based on various generation strategies and adapt naturally through social network evolution processes \cite{mcpherson2001birds}\cite{daud2020applications}.
Once the HASNs are generated using one of the proposed strategies, the social network is expected to evolve over time, potentially forming new connections among individuals. To model this evolution, we employ link prediction, a widely used technique for identifying potential connections between previously unconnected nodes in social networks. One classical approach to link prediction is the Jaccard similarity index \cite{arrar2024comprehensive}\cite{daud2020applications}, which quantifies the similarity (or likelihood) between two nodes by calculating the ratio of the intersection to the union of their immediate neighbors. Mathematically, the Jaccard similarity is defined as: $\sigma_{Jaccard}(x,y) = \frac{|N(x) \cap N(y)|}{|N(x) \cup N(y)|}$, where $N(x)$ indicates the number of neighbors of node $x$. By computing the Jaccard similarity for all pairs of nodes, we can select the top $k$ pairs with the highest similarity values to establish new connections. This process is then iterated over $r$ rounds to simulate the evolution of the network.

\section{Supplementary Experimental Results}
\label{sec: Supplementary Experimental Results}

\subsection{Basic Statistics of Network Datasets}
\label{sec: Basic Statistics of Network Datasets}
Table \ref{dataset statistics} displays basic statistics for the three datasets used in our experiments, while Table \ref{Basic statistics of datasets before and after evolution} provides the statistics of these datasets after inserting AI nodes using the proposed generation strategies, along with the statistics of the AI node-augmented networks before and after evolution (\textit{cf.} Appendix \ref{sec: The Proposed HASN Scenarios}).

\begingroup
\renewcommand{\arraystretch}{1}
\begin{table}[t]
\centering
\caption{Basic statistics of benchmark datasets.}
\label{dataset statistics}
\resizebox{\columnwidth}{!}{%
\begin{tabular}{lcccc}
\toprule
Dataset & \#Nodes & \#Edges & Avg. Degree\\
\midrule
Cora \cite{sen2008collective} & 2708 & 5429 & 4.01 \\
CiteSeer \cite{sen2008collective} & 3327 & 4732 & 2.84 \\
PubMed \cite{sen2008collective} & 19717 & 44338 & 4.50 \\
\bottomrule
\end{tabular}%
}
\end{table}
\endgroup

\begin{table}[t]
\centering
\caption{The Q and HQ results on the Cora dataset obtained by CUSA under different hh, ha, and aa settings.}
\label{hh, ha, and aa settings}
\LARGE
\begin{tabular}{lcc}
\toprule
Setting & Q & HQ \\
\midrule
$hh,ha,aa=1,1,1$ & 0.8215 & 0.8200 \\
$hh,ha,aa=3,2,1$ & 0.8228 & 0.8216 \\
$hh,ha,aa=4,2,1$ & \textbf{0.8230} & \textbf{0.8224} \\
$hh,ha,aa=5,3,1$ & 0.8226 & 0.8214 \\
$hh,ha,aa=1,2,3$ & 0.8204 & 0.8193 \\
$hh,ha,aa=1,3,5$ & 0.8201 & 0.8190 \\
$hh,ha,aa=1,5,1$ & 0.8218 & 0.8212 \\
$hh,ha,aa=1,5,3$ & 0.8214 & 0.8210 \\
$hh,ha,aa=3,5,1$ & 0.8208 & 0.8204 \\
\bottomrule
\end{tabular}
\end{table}

\subsection{The Sensitivity to \textit{hh,ha,aa}}
\label{Sensitivity to hh, ha, aa}
Following \cite{zhang2022new}\cite{de2021centrality}, the values of $hh$, $ha$, and $aa$ are positively correlated with the importance of edge types. Since MetaCD emphasizes human-centric community detection, edges connecting human nodes are assigned higher weights, leading to $hh > ha > aa$. While these settings serve as the default in our experiments, they are flexible and can be adjusted as needed.

Table \ref{hh, ha, and aa settings} compares the modularity Q and human-centric modularity HQ results obtained by CUSA under different settings of $hh$, $ha$, and $aa$ on the Cora dataset, where 20\% of the total nodes are inserted as AI nodes using $k\%$ random insertion. Under the configuration $hh > ha > aa$, the values of Q and HQ obtained by CUSA consistently exceed those achieved under other configurations, particularly outperforming the $aa > ha > hh$ setting. This demonstrates that prioritizing human-human edges during AI scoring improves the evaluation of AI nodes in fostering human interactions, aligning with our goal of integrating AI nodes within communities. Furthermore, we observe that CUSA is sensitive only to the relative magnitudes of $hh$, $ha$, and $aa$, rather than their absolute values. As a result, setting these weights in practical applications becomes relatively straightforward.

\begingroup
\renewcommand{\arraystretch}{1.08}
\begin{table}[t]
\centering
\caption{The Q and HQ results on the Cora dataset obtained by CUSA under different inserted ratios of AI nodes.}
\label{different ai ratio}
\resizebox{\columnwidth}{!}{%
\begin{tabular}{cccc}
\toprule
AI node ratio & \#Nodes, \#Edges & Q & HQ \\
\midrule
0\% & (2708, 5429) & 0.8154 & 0.8154 \\
10\% & (2978, 6238) & 0.8314 & 0.8242 \\
20\% & (3249, 6896) & 0.8343 & 0.8236 \\
30\% & (3520, 7355) & 0.8358 & 0.8253 \\
50\% & (4062, 8723) & 0.8411 & 0.8337 \\
100\% & (5416, 12677) & 0.8478 & 0.8396 \\
\bottomrule
\end{tabular}%
}
\end{table}
\endgroup

\begingroup
\renewcommand{\arraystretch}{1.2}
\begin{table}[t]
\centering
\caption{The execution time (minutes) on two large-scale datasets \cite{yang2012defining} to assess the scalability of CUSA.}
\label{Scalability of CUSA}
\resizebox{\columnwidth}{!}{%
\begin{tabular}{lcccccc}
\toprule
\textbf{\#Edges} & 10K & 100K & 200k & 300K & 500K & 1M \\
\toprule
YouTube & 35K & 684K & 1.1M & 1.5M & 2M & 2.8M \\
LiveJournal & 45K & 1.1M & 2.1M & 3.9M & 8.2M & 140M \\
\toprule
\textbf{Execution time} \\
\toprule
YouTube & 11.1 & 27.51 & 44.96 & 70.28 & 75.95 & 121.88 \\
LiveJournal & 16.88 & 55.50 & 80.83 & 136.93 & 298.95 & 646.51 \\
\bottomrule
\end{tabular}%
}
\end{table}
\endgroup

\subsection{The Sensitivity to Different AI Ratios}
\label{Sensitivity to Different AI Ratios}
Table \ref{different ai ratio} presents the Q and HQ results obtained by CUSA (with $hh, ha, aa = 4,2,1$) on the Cora dataset with varying inserted proportions of AI nodes (i.e., $\frac{\text{\# AI nodes}}{\text{\# total nodes}}$), using the third generation strategy (i.e., introverted and extroverted AI equally distributed). As more AI nodes are inserted, Q increases, indicating that the newly added AI nodes effectively facilitate social interactions and contribute to forming more tightly-knit communities. More importantly, CUSA maintains high HQ with a large number AI nodes inserted, demonstrating that CUSA can select an appropriate number of AI nodes for community detection to enhance modularity while avoiding an excessive number of AI nodes within communities.

\subsection{The Scalability of CUSA}

Table \ref{Scalability of CUSA} evaluates the scalability of CUSA using two large-scale datasets, YouTube \cite{yang2012defining} and LiveJournal \cite{yang2012defining}. The upper part of Table \ref{Scalability of CUSA} (above the third line) presents the number of edges within the extracted subgraphs for varying numbers of nodes (i.e., 10K, 100K, ..., 1M) across both datasets, while the lower part of Table \ref{Scalability of CUSA} displays the execution time (in minutes) of CUSA on these subgraphs. Although the theoretical time complexity is $O(|AI|(|V||E| + |V|^2 \log |V|))$, the observed execution time grows approximately linearly with $|E|$, as $|E|$ dominates the overall complexity. This demonstrates that in real-world large-scale social networks, CUSA achieves practical scalability closer to $O(|E|)$. 

\begingroup
\renewcommand{\arraystretch}{1.02}
\begin{table*}[t]
\centering
\caption{Basic statistics of AI node-augmented networks before and after evolution, where "Before Evolution" (BE) and "After Evolution" (AE) are indicated.}
\label{Basic statistics of datasets before and after evolution}
\resizebox{\textwidth}{!}{%
\begin{tabular}{lccccc}
\toprule
\textbf{(1) \textit{k\%} random insertion} \\
\midrule
Dataset & \#Nodes & \#Edges (BE) & Avg. Degree (BE) & \#Edges (AE) & Avg. Degree (AE) \\
\midrule
Cora & 2978 & 5665 (+4.3\%) & 3.80 & 6065 (+11.7\%) & 4.07 \\
CiteSeer & 3659 & 5198 (+9.8\%) & 2.84 & 5598 (+18.3\%) & 3.06 \\
PubMed & 21688 & 46581 (+5.0\%) & 4.29 & 48981 (+10.5\%) & 4.52 \\
\midrule
\textbf{(2) Inverse degree-based insertion}  \\
\midrule
Dataset & \#Nodes & \#Edges (BE) & Avg. Degree (BE) & \#Edges (AE) & Avg. Degree (AE) \\
\midrule
Cora & 2978 & 5680 (+4.6\%) & 3.81 & 6080 (+12.0\%) & 4.08 \\
CiteSeer & 3659 & 5125 (+8.3\%) & 2.80 & 5525 (+16.8\%) & 3.02 \\
PubMed & 21688 & 46352 (+4.5\%) & 4.27 & 48754 (+10.0\%) & 4.50 \\
\midrule
\textbf{(3) Inner and outer AI mix}  \\
\midrule
Dataset & \#Nodes & \#Edges (BE) & Avg. Degree (BE) & \#Edges (AE) & Avg. Degree (AE) \\
\midrule
Cora & 2978 & 5838 (+7.5\%) & 3.92 & 6238 (+14.9\%) & 4.19 \\
CiteSeer & 3659 & 5037 (+6.4\%) & 2.75 & 5437 (+14.9\%) & 2.97 \\
PubMed & 21688 & 48520 (+9.4\%) & 4.47 & 50920 (+14.8\%) & 4.69 \\
\midrule
\textbf{(4) AI with dual personality}  \\
\midrule
Dataset & \#Nodes & \#Edges (BE) & Avg. Degree (BE) & \#Edges (AE) & Avg. Degree (AE) \\
\midrule
Cora & 2978 & 5862 (+8.0\%) & 3.94 & 6262 (+15.3\%) & 4.21 \\
CiteSeer & 3659 & 5258 (+11.1\%) & 2.84 & 5658 (+19.6\%) & 3.09 \\
PubMed & 21688 & 46074 (+3.9\%) & 4.25 & 48474 (+9.3\%) & 4.47 \\
\bottomrule
\end{tabular}%
}
\end{table*}
\endgroup

\begingroup
\renewcommand{\arraystretch}{1.2}
\begin{table*}[t]
\centering
\caption{The Q and HQ results on the Cora dataset transformed into HASNs  using different generation strategies (without evolving), obtained by CUSA in comparison to that of four clustering baselines each equipped with no removal of AIs (N), all removal of AIs (A), and removal of AIs by GAD (D).}
\label{The Q and HQ results on the Cora dataset (without evolving)}
\resizebox{\textwidth}{!}{%
\begin{tabular}{cccccccccc}
\toprule
\multirow{2}{*}{\makebox[0.15\textwidth]{\textbf{Method}}} & \multirow{2}{*}{} & \multicolumn{2}{c}{\textbf{\textit{k\%} random insertion}} & \multicolumn{2}{c}{\textbf{Inverse degree insertion}} & \multicolumn{2}{c}{\textbf{Inner and outer AI mix}} & \multicolumn{2}{c}{\textbf{AI with dual personality}} \\
\cline{3-10}
& & \textbf{Q (↑)} & \textbf{HQ (↑)} & \textbf{Q (↑)} & \textbf{HQ (↑)} & \textbf{Q (↑)} & \textbf{HQ (↑)} & \textbf{Q (↑)} & \textbf{HQ (↑)} \\
\hline
\multirow{3}{*}{\makebox[0.15\textwidth]{Spectral \cite{amini2013pseudo}}} 
& N & 0.1942 & 0.1847 & 0.2387 & 0.2270 & 0.2718 & 0.2588 & 0.1744 & 0.1664 \\
& A & 0.2890 & 0.2890 & 0.2890 & 0.2890 & 0.2890 & 0.2890 & 0.2890 & 0.2890 \\
& D & 0.2520 & 0.2520 & 0.2653 & 0.2653 & 0.2687 & 0.2687 & 0.2662 & 0.2612 \\
\hline
\multirow{3}{*}{\makebox[0.15\textwidth]{Louvain \cite{blondel2008fast}}} 
& N & 0.7954  & 0.7353 & 0.7973 & 0.7428 & 0.8183 & 0.7973 & 0.7738 & 0.7317 \\
& A & 0.8112 & 0.8112 & 0.8112 & 0.8112 & 0.8112 & 0.8112 & 0.8112 & 0.8112 \\
& D & 0.8027 & 0.7829 & 0.7957 & 0.7680 & 0.8110 & 0.8068 & 0.7968 & 0.7886 \\
\hline
\multirow{3}{*}{\makebox[0.15\textwidth]{GCC \cite{fettal2022efficient}}} 
& N & 0.6862 & 0.6451 & 0.6925 & 0.6658 & 0.7052 & 0.6787 & 0.6942 & 0.6614 \\
& A & 0.7323 & 0.7323 & 0.7323 & 0.7323 & 0.7323 & 0.7323 & 0.7323 & 0.7323 \\
& D & 0.7131 & 0.7042 & 0.7218 & 0.7139 & 0.7286 & 0.7220 & 0.7164 & 0.7035 \\
\hline
\multirow{3}{*}{\makebox[0.15\textwidth]{LSEnet \cite{sun2024lsenet}}} 
& N & 0.7026 & 0.6543 & 0.7073 & 0.6719 & 0.7145 & 0.7020 & 0.7011 & 0.6756 \\
& A & 0.7258 & 0.7258 & 0.7258 & 0.7258 & 0.7258 & 0.7258 & 0.7258 & 0.7258 \\
& D & 0.7064 & 0.6915 & 0.7165 & 0.7093 & 0.7192 & 0.7130 & 0.7181 & 0.7132 \\
\hline
\multirow{1}{*}{\makebox[0.15\textwidth]{\textbf{CUSA} (ours)}} 
& - & \textbf{0.8141} & \textbf{0.8132} & \textbf{0.8166} & \textbf{0.8160} & \textbf{0.8236} & \textbf{0.8137} & \textbf{0.8124} & \textbf{0.8120} \\
\bottomrule
\end{tabular}%
}
\end{table*}
\endgroup

\begin{table}[t]
\centering
\caption{The Q and HQ results on the ER-based random HASN \cite{erdds1959random} (with evolution), obtained by CUSA in comparison to that of four clustering baselines each equipped with no removal of AIs (N), all removal of AIs (A), and removal of AIs by GAD (D).}
\label{The Q and HQ results on the ER-based random HASN}
\resizebox{\columnwidth}{!}{%
\begin{tabular}{ccccc}
\toprule
\multirow{2}{*}{\makebox[0.15\textwidth]{\textbf{Method}}} & \multirow{2}{*}{} & \multicolumn{2}{c}{\textbf{\textit{k\%} random insertion}} \\
\cline{3-4}
& & \textbf{Q (↑)} & \textbf{HQ (↑)} \\
\hline
\multirow{3}{*}{\makebox[0.15\textwidth]{Spectral \cite{amini2013pseudo}}} 
& N & 0.0106 & 0.0088 \\
& A & 0.0056 & 0.0056 \\
& D & 0.0136 & 0.0131 \\
\hline
\multirow{3}{*}{\makebox[0.15\textwidth]{Louvain \cite{blondel2008fast}}} 
& N & 0.4791  & 0.4461 \\
& A & 0.4923 & 0.4923 \\
& D & 0.4882 & 0.4792 \\
\hline
\multirow{3}{*}{\makebox[0.15\textwidth]{GCC \cite{fettal2022efficient}}} 
& N & 0.3425 & 0.3128 \\
& A & 0.3743 & 0.3743 \\
& D & 0.3675 & 0.3597 \\
\hline
\multirow{3}{*}{\makebox[0.15\textwidth]{LSEnet \cite{sun2024lsenet}}} 
& N & 0.3732 & 0.3569 \\
& A & 0.4215 & 0.4135 \\
& D & 0.3968 & 0.3913 \\
\hline
\multirow{1}{*}{\makebox[0.15\textwidth]{\textbf{CUSA} (ours)}} 
& - & \textbf{0.4993} & \textbf{0.4972} \\
\bottomrule
\end{tabular}%
}
\end{table}

\subsection{Experiments on HASNs without Evolution}
Table \ref{The Q and HQ results on the Cora dataset (without evolving)} assesses Q and HQ results of all methods on the Cora dataset transformed into HASNs using the four generation strategies, without evolving. For consistency, the other settings align with Table \ref{The Q and HQ results on the Cora dataset}. Compared to Table \ref{The Q and HQ results on the Cora dataset}, the performance trends of each method across different generation strategies remain largely unchanged, though the values of Q and HQ show a slight decline. This is because AI nodes, when initially inserted into the network, only influence the topology of their immediate neighbors, limiting their ability to facilitate social interactions across the entire network. To better reflect real-world scenarios, where AI nodes foster the formation of new social connections, we incorporate social network evolution in our experiments following the insertion of AI nodes (\textit{cf.} Appendix \ref{sec: The Evolution of HASNs}).

\subsection{Experiments on Random Networks}
In Table \ref{The Q and HQ results on the ER-based random HASN}, we conduct experiments on an Erdős-Rényi-based random HASN (ER-based HASN) \cite{erdds1959random} aligning with the Cora dataset and demonstrate that CUSA effectively distinguishes the community structures in real-world networks from those in random networks. Following \cite{erdds1959random}, we generate a random network with 2708 nodes and 5429 edges, aligning with Cora dataset. The ER-based HASN is then constructed by inserting AI nodes (10\% of the total nodes) using k\% random insertion, followed by evolution, resulting in a network with 2978 nodes and 6254 edges. We evaluate Q and HQ across all methods on the ER-based random HASN, as shown in Table \ref{The Q and HQ results on the ER-based random HASN}. Compared to the results on Cora presented in Table \ref{The Q and HQ results on the Cora dataset}), both Q and HQ values are significantly lower, reflecting the fundamental differences between real-world and random networks. Since the ER-based HASN inherently lack community structures, its lower Q scores confirm that CUSA successfully detects meaningful communities in structured datasets such as Cora, where it obtains Q = 0.8190. In contrast, the ER-based HASN exhibits a much lower Q = 0.4993, further validating the ability of CUSA to distinguish real-world community structures from randomly generated networks.

\begin{table}[t]
\centering
\caption{Statistics of Timik (3,000-node subgraph)}
\label{timik_statistics}
\resizebox{\columnwidth}{!}{%
\begin{tabular}{ccccccc}
\toprule
Dataset & \#N & \#E & \#AIs & AD & H-AI AD & H-H AD \\
\midrule
Timik & 3000 & 9258 & 1396 & 6.17 & 3.75 & 2.39 \\
\bottomrule
\end{tabular}
}
\end{table}

\begingroup
\renewcommand{\arraystretch}{1.152}
\begin{table}[t]
\centering
\caption{The basic statistical analysis of the real-world HASN (Timik) \cite{chen2022user}, where InRatio denotes the proportion of intra-community edges for each AI node.}
\label{The basic statistical analysis of the real-world HASN}
\resizebox{\columnwidth}{!}{%
\begin{tabular}{lcc}
\toprule
Interval of InRatio & \#AI nodes & Proportion of Total AIs \\
\midrule
$[0.0, 0.1)$ & 480 & 34.38\% \\
$[0.1, 0.2)$ & 1 & 0.07\% \\
$[0.2, 0.3)$ & 11 & 0.79\% \\
$[0.3, 0.4)$ & 26 & 1.86\% \\
$[0.4, 0.5)$ & 16 & 1.15\% \\
$[0.5, 0.6)$ & 134 & 9.60\% \\
$[0.6, 0.7)$ & 82 & 5.87\% \\
$[0.7, 0.8)$ & 52 & 3.72\% \\
$[0.8, 0.9)$ & 46 & 3.30\% \\
$[0.9, 1.0]$ & 548 & 39.26\% \\
\bottomrule
\end{tabular}
}
\end{table}
\endgroup

\subsection{The Validation of Generation Strategies}
\label{Validation of Generation Strategies}
In Table \ref{timik_statistics} and Table \ref{The basic statistical analysis of the real-world HASN}, we validate the proposed generation strategies, particularly strategies (2)–(4) (i.e., inverse degree-based insertion, introverted and extroverted AI distribution, and AI with dual personality) (\textit{cf.} Appendix \ref{sec: The Proposed HASN Scenarios}), by analyzing statistics from a real-world HASN, Timik \cite{chen2022user}. Based on AI interaction patterns, we categorize our generation strategies into two groups: (1) Personality-oriented: This includes introverted and extroverted AI distribution, as well as AI with dual personality. AI nodes are designed to exhibit human-like personality traits, ranging from introverted to extroverted or a combination of both \cite{zhang2024better}. (2) Task-oriented: This involves inverse degree-based insertion, where AI nodes preferentially connect to low-degree human nodes to help reduce loneliness \cite{skjuve2021my}\cite{loveys2019reducing}.

Timik is a dataset collected from a social metaverse, where digital twins replace offline users and continue their activities in the virtual world \cite{chen2022user}. In this study, we treat digital twins as AI nodes. Since the original Timik network is significantly larger, we extract a 3,000-node subgraph for analysis while preserving the structural properties. The basic statistics of this Timik subgraph are shown in Table \ref{timik_statistics}, where AD stands for the "Average Degree". To analyze AI behavior patterns in Timik, we conduct basic statistical analyses on its AI nodes to examine two types of interaction patterns:

\noindent \textbf{(1) Personality-oriented patterns}: To determine whether AI nodes follow personality-oriented patterns, we calculated the proportion of intra-community edges for each AI node (denoted by InRatio). A higher InRatio indicates an introverted AI, while a lower InRatio suggests an extroverted AI. An InRatio around 0.5 implies dual-personality AI. The corresponding results are presented in Table \ref{The basic statistical analysis of the real-world HASN}.  Inspection of Table \ref{The basic statistical analysis of the real-world HASN}, we observe that nearly 40\% of AI nodes are introverted, another 40\% are extroverted, and approximately 10\% possess a dual personality. This validates that the introverted and extroverted AI equally distributed generation strategy closely resembles the patterns found in Timik. Besides, although the AI with dual personality strategy differs slightly from Timik, the presence of dual AI in Timik is relatively common, supporting its relevance. 

\noindent \textbf{(2) Task-oriented patterns}: To examine whether AI nodes follow task-oriented patterns, we compared the average number of human friends between all human nodes and those connected to AI nodes.  As shown in Table \ref{timik_statistics}, the Human-AI and Human-Human average degrees (denoted by H-AI AD and H-H AD) are 3.75 and 2.39, respectively. Since human nodes connected to AI nodes do not exhibit significantly fewer human friends, the inverse degree-based insertion strategy does not align with the AI behavior patterns observed in Timik. This discrepancy may arise from the fact that AI nodes in Timik, acting as digital twins, mimic the behavior of their original users, most of whom do not intentionally aim to reduce others' loneliness.

\begingroup
\renewcommand{\arraystretch}{1}
\begin{table}[t]
\centering
\caption{The Q and HQ results on the real-world dataset (Timik) \cite{chen2022user}, obtained by CUSA in comparison to that of four clustering baselines each equipped with no removal of AIs (N), all removal of AIs (A), and removal of AIs by GAD (D).}
\label{The Q and HQ results on the Timik dataset}
\small
\resizebox{\columnwidth}{!}{%
\begin{tabular}{cccc}
\toprule
\textbf{Method} & & \textbf{Q (↑)} & \textbf{HQ (↑)} \\
\midrule
\multirow{3}{*}{\makebox[0.15\textwidth]{Spectral \cite{amini2013pseudo}}} 
& N & 0.0275 & 0.0146 \\
& A & 0.0198 & 0.0198 \\
& D & 0.0232 & 0.0152 \\
\midrule
\multirow{3}{*}{\makebox[0.15\textwidth]{Louvain \cite{blondel2008fast}}} 
& N & 0.4038  & 0.2120 \\
& A & 0.3721 & 0.3721 \\
& D & 0.3812 & 0.2638 \\
\midrule
\multirow{3}{*}{\makebox[0.15\textwidth]{GCC \cite{fettal2022efficient}}} 
& N & 0.3653 & 0.1921 \\
& A & 0.3437 & 0.3437 \\
& D & 0.3512 & 0.2154 \\
\midrule
\multirow{3}{*}{\makebox[0.15\textwidth]{LSEnet \cite{sun2024lsenet}}} 
& N & 0.3712 & 0.2021 \\
& A & 0.3540 & 0.3540 \\
& D & 0.3658 & 0.2787 \\
\midrule
\multirow{1}{*}{\makebox[0.15\textwidth]{\textbf{CUSA} (ours)}} 
& - & \textbf{0.4324} & \textbf{0.4238} \\
\bottomrule
\end{tabular}%
}
\end{table}
\endgroup

\subsection{Experimental Results on Real-world HASNs}
In Table \ref{The Q and HQ results on the Timik dataset}, we carry out experiments on Timik, a real-world HASN, to evaluate the practical utility of CUSA in comparison to all baseline methods. The corresponding results are presented in Table \ref{The Q and HQ results on the Timik dataset}, from which two key observations can be drawn. 

First, CUSA outperforms all baselines across all removal strategies in terms of both Q and HQ. Even with AI nodes comprising 46\% of the total nodes in Timik, CUSA achieves high performance by leveraging AI Scoring to carefully select an appropriate number of AI nodes (ensuring a high human-to-AI ratio) to participate in communities while preserving high modularity. In contrast, the all-removal strategy (A) results in the lowest Q, as completely disregarding AI nodes creates a loosely connected structure in Timik, making it difficult to identify well-defined community structures due to sparse interconnections. The no-removal strategy (N) achieves a high Q but a significantly low HQ due to an extremely low human-to-AI ratio. Similarly, the GAD-based AI removal strategy (G) yields a low HQ relative to Q because most AI nodes remain embedded within communities. This highlights the limitations of GAD in distinguishing influential AI nodes from others, particularly when AI nodes exhibit human-like behaviors in real-world applications, ultimately leading to poor human-to-AI ratios and lower HQ. 

Second, it is noteworthy that CUSA consistently identifies communities with Q and HQ across varying proportions of AI nodes. This is achieved by carefully assessing each AI node's impact on network interactions through AI Scoring. By contrast, the all-removal strategy (A) is highly sensitive to AI node proportions. It performs well at low AI ratios (e.g., 10\%, as shown in Tables \ref{AI node scoring methods}, \ref{The HQ results on different datasets}, and \ref{The Q and HQ results on the Cora dataset}) but declines sharply as AI presence increases, as seen in Timik, where the all-removal strategy lags far behind CUSA. This highlights the limitations of indiscriminate AI removal and underscores the need for AI-aware clustering to balance human-centric communities while adapting to AI’s evolving role in social networks.

\begingroup
\renewcommand{\arraystretch}{1.1}
\begin{table}[t]
\centering
\caption{The HMR and ADM results on different datasets obtained by CUSA.}
\label{The HMR and ADM results of CUSA across different datasets}
\resizebox{\columnwidth}{!}{%
\begin{tabular}{lccc}
\toprule
Dataset & Avg. Degree & HMR (\%) & ADM \\
\midrule
Cora & 4.01 & 14.09\% & 20.41 \\
CiteSeer & 2.84 & 56.41\% & 56.3 \\
PubMed & 4.50 & 42.03\% & 435.74 \\
Timik & 6.17 & 30.47\% & 47.71 \\
\bottomrule
\end{tabular}
}
\end{table}
\endgroup

\subsection{The HMR and ADM Results on Different Datasets} 
\label{HMR and ADM Results on Different Datasets}
Table \ref{The HMR and ADM results of CUSA across different datasets} demonstrates the HMR and ADM results of CUSA on different datasets. The findings indicate that incorporating AI nodes in community detection reshapes the identified communities, influencing not just immediate neighbors but also broader network structures. This is evident from ADM values, which are significantly higher than the average degree, underscoring the substantial role of AI nodes in community formation.

\end{document}